\documentclass{article}     
\usepackage[dvips]{graphics}     
\usepackage{eepic}     
\newcommand{\sgn}{\mathrm{sgn}}     
\newcommand{\degr}{^{\circ}}     
     
\begin{document}     
     
\title{\bf       
Two--Frequency Forced Faraday Waves:\\ Weakly Damped Modes and     
Pattern Selection\footnote{To appear in a special issue of \textbf{Physica D}
dedicated to the memory of John David Crawford.}}     
      
\author{Mary Silber and Chad M.~Topaz\\      
{\it Department of Engineering Sciences \& Applied Mathematics}\\      
{\it Northwestern University} \\      
{\it Evanston, IL 60208 USA}      
\and {Anne C.~Skeldon}\\       
{\it Department of Mathematics}\\       
{\it City University, Northampton Square}\\       
{\it London, EC1V OHB, UK}}      
\date{April 17, 2000}     
     
\maketitle      
\Large{\centerline{\bf To the memory of John David Crawford}}      
\normalsize
      
\begin{abstract}      
 
Recent experiments~\cite{ref:kpg98} on two--frequency parametrically
excited surface waves produce an intriguing ``superlattice'' wave
pattern near a codimension--two bifurcation point where both
subharmonic and harmonic waves onset simultaneously, but with
different spatial wavenumbers.  The superlattice pattern is
synchronous with the forcing, spatially periodic on a large hexagonal
lattice, and exhibits small--scale triangular structure.  Similar
patterns have been shown to exist as primary solution branches of a
generic 12--dimensional ${\rm D}_6\dot{+}{\rm T}^2$--equivariant
bifurcation problem, and may be stable if the nonlinear coefficients
of the bifurcation problem satisfy certain
inequalities~\cite{ref:sp98}. Here we use the spatial and temporal
symmetries of the problem to argue that weakly damped harmonic waves
may be critical to understanding the stabilization of this pattern in
the Faraday system.  We illustrate this mechanism by considering the
equations developed by Zhang and Vi\~{n}als~\cite{ref:zv97a} for small
amplitude, weakly damped surface waves on a semi--infinite fluid
layer. We compute the relevant nonlinear coefficients in the
bifurcation equations describing the onset of patterns for excitation
frequency ratios of 2/3 and 6/7.  For the 2/3 case, we show that there
is a fundamental difference in the pattern selection problems for
subharmonic and harmonic instabilities near the codimension--two point.
Also, we find that the 6/7 case is significantly different from the
2/3 case due to the presence of additional weakly damped harmonic
modes.  These additional harmonic modes can result in a stabilization
of the superpatterns.
 
\end{abstract}      
      
\section{Introduction}       
\label{sec:introduction}       

Faraday waves are parametrically excited on the free surface of a
fluid layer when it is subjected to a vertical vibration of sufficient
strength. This pattern--forming hydrodynamic system has proven to be
an especially versatile one in laboratory
experiments~\cite{ref:bemj95,ref:kg96}, exhibiting the common patterns
familiar from convection (stripes, squares, hexagons, spirals), as
well as more exotic patterns such as triangles~\cite{ref:m93},
quasipatterns~\cite{ref:kpg98,ref:ef94,ref:cal92}, superlattice
patterns~\cite{ref:kpg98,ref:wmk99,ref:af98}, time--dependent rhombic
patterns~\cite{ref:afpreprint} and localized
waves~\cite{ref:wmk99,ref:af00}. See~\cite{ref:mfp98} for a recent
review paper on Faraday wave pattern formation.

The temporal period of the Faraday waves is typically twice that of the
vibration in the case of purely sinusoidal forcing.  The observation
of this subharmonic response is attributed to Faraday~\cite{ref:f1831}
and was first explained theoretically by Benjamin and Ursell's linear
stability analysis for inviscid, potential flow~\cite{ref:bu54}. More
recently it has been shown that waves, {\it synchronous} with the forcing,
can be excited in thin layers of fluid vibrated at low frequency
~\cite{ref:k96,ref:mwwak97,ref:ct97}; in certain viscoelastic
fluids~\cite{ref:wmk99}; and in fluids forced periodically, but with
more than one frequency component~\cite{ref:ef94,ref:bf95,ref:bet96}.
In each of these Faraday systems it is possible to tune the forcing
parameters in order to access the transition between subharmonic and
harmonic response. At this codimension--two point, both instabilities
set in simultaneously, but with different spatial wavenumbers.
  
Many of the
experimental~\cite{ref:m93,ref:ef94,ref:af98,ref:afpreprint,ref:af00,ref:bv97,ref:bwv97}
and theoretical studies~\cite{ref:zv97a,ref:zv97b,ref:lp97,ref:cv99}
of exotic patterns in the Faraday system attribute their formation
near the codimension--two (or ``bicritical'') point to resonant triad
interactions involving the critical or near--critical modes with
different spatial wavenumbers. In particular, the focus has been on
spatial triads ${\bf k}_1$, ${\bf k_2}$ and ${\bf k}_3={\bf
k}_1\pm{\bf k}_2$, where $|{\bf k}_1|=|{\bf k}_2|$ is the wavenumber
of one critical mode, and $|{\bf k}_3|$ is the wavenumber of the other
critical mode.  The angle $\theta_r$, which separates ${\bf k}_1$ and
${\bf k}_2$, is readily tuned by changing the frequency components
$m\omega$ and $n\omega$ of a two--frequency periodic forcing
function. It has been suggested, for example, that by tuning this
angle, different types of exotic wave patterns may be
selected~\cite{ref:ef94}.  Such a simple mechanism for {\it nonlinear}
pattern selection, which is based on examining the {\it linear}
instabilities of the spatially homogeneous state, is naturally
attractive, but warrants careful examination as we show.

Silber and Skeldon~\cite{ref:ss99} recently showed that whether or not
resonant triads associated with the bicritical point affect pattern
selection depends on the temporal characteristics of the competing
instabilities.  For instance, the bicritical point of laboratory
experiments typically involves a subharmonic mode (Floquet multiplier
$-1$) and a harmonic mode (Floquet multiplier $+1$).  On the
subharmonic side of the bicritical point, the onset pattern selection
problem is strongly influenced by the presence of the weakly damped
harmonic mode. In contrast, on the harmonic side, the onset pattern
selection problem is completely insensitive to the presence of near
critical subharmonic modes.  These general ideas were demonstrated
in~\cite{ref:ss99} through a bifurcation analysis of a hydrodynamic
model of one--dimensional Faraday waves.

Here, we extend the bifurcation analysis in~\cite{ref:ss99} to
two--dimensional spatially--periodic patterns and to higher forcing
frequencies within the two--frequency forcing function. With the
experimentally--relevant higher forcing frequencies ({\it e.g.}
$6\omega$ and $7\omega$) employed in this paper, we find the new
possibility that spatially--resonant triads involving nearly critical
{\it harmonic} modes may influence the {\it harmonic} wave pattern
selection problem. This is not an option for the lower forcing
frequencies ({\it e.g.} $1\omega/2\omega$ and $2\omega/3\omega$) used
in previous weakly nonlinear analyses of the two--frequency Faraday
problem~\cite{ref:zv97b,ref:ss99}.

We follow J.D.~Crawford's seminal work on Faraday
waves~\cite{ref:c90a,ref:c91c,ref:c91b,ref:cgl93} by posing the
pattern selection problem in terms of a symmetry--breaking bifurcation
of the trivial fixed--point of a stroboscopic map. By restricting
solutions to those that are spatially--periodic on some hexagonal
lattice we obtain a finite--dimensional bifurcation problem that can
be analyzed using the methods of equivariant bifurcation
theory~\cite{ref:gss}. For a review of this approach to hydrodynamic
pattern formation problems, see Crawford and Knobloch~\cite{ref:ck91}.  
  
This formulation of the bifurcation problem allows us to address recent
two--frequency Faraday wave experimental observations~\cite{ref:kpg98}
of a transition between simple hexagons and the triangular
superlattice wave pattern depicted in
Figure~\ref{fig:67sl}a. Specifically, we follow~\cite{ref:sp98} and
consider a bifurcation problem that is equivariant with respect to a
twelve--dimensional irreducible representation of ${\rm D}_6\dot{+}
{\rm T}^2$, which is analyzed in~\cite{ref:dg92,ref:dss97}. The
observed harmonic wave states correspond to primary transcritical
branches of the generic bifurcation problem. In order for the observed
hexagon--superlattice pattern transition to be reproduced by the
bifurcation problem, we must consider a degenerate case in which the
quadratic coefficient vanishes. Moreover, the cubic coefficients must
satisfy certain inequalities, {\it e.g.} certain combinations of
nonlinear cross--coupling coefficients must be small compared to the
cubic self--coupling coefficient.

In this paper we compute the quadratic and cubic nonlinear
coefficients in the bifurcation problem from the Zhang--Vi\~{n}als
equations~\cite{ref:zv97b} which apply to deep layers of low viscosity
fluids subjected to a periodic acceleration. We show that the
necessary inequalities for stable superlattice patterns can be
satisfied for the forcing frequencies employed in the experiments
($6\omega/7\omega$), and that a resonant triad involving a weakly
damped harmonic mode plays a key role in stabilizing the
superpattern. Specifically, we find that the presence of a near
critical harmonic mode leads to a cancellation in one of the cubic
cross--coupling coefficients, causing this coefficient to become small
in magnitude as required. This selects a preferred angle $\theta_r$
for the superlattice patterns. In other words, it suggests which of
the countably infinite 12--dimensional irreducible representations of
${\rm D}_6\dot{+} {\rm T}^2$ is most pertinent to this Faraday wave
problem.

The paper is organized as follows.  Section \ref{sec:linear} presents
background linear stability results for the two--frequency Faraday
experiment, while section \ref{sec:resonants} reviews results from
\cite{ref:ss99} on the influence of spatio--temporally resonant triads
on pattern selection.  Section~\ref{sec:bifurcation} then formulates
the generic bifurcation problem relevant to our investigation. The
bifurcation results derived from the two--frequency Faraday problem
modeled by the Zhang--Vi\~{n}als equations are presented in
Section~\ref{sec:results}; the coefficients of the leading nonlinear
terms are evaluated numerically from expressions derived
perturbatively in the Appendix.  We consider two different cases. In
Section~\ref{sec:res23} we consider an example involving forcing
frequencies in ratio $m/n=2/3$, focusing on differences between the
pattern selection problems for subharmonic and harmonic wave onset in
a vicinity of the bicritical point.  Section~\ref{sec:res67} then
turns to an example involving higher forcing frequencies in ratio
$m/n=6/7$, and shows how weakly damped harmonic modes can stabilize
harmonic wave superpatterns involving the angle $\theta_r$ associated
with a harmonic wave resonant triad. Finally,
Section~\ref{sec:conclude} concludes the paper with a brief summary of
our results and some discussion of issues we hope to address in the
future.

\section{Background}      
\label{sec:background}      
      
\subsection{Linear Results}     
\label{sec:linear}     
     
In the two-frequency Faraday wave problem a container of fluid is 
accelerated in the vertical direction with an excitation of the form 
\begin{equation}     
\label{eq:f(t)}     
g(t)=g_0+g_z\bigl( \cos (\chi) \cos (m\omega t) + \sin (\chi)      
\cos (n \omega t + \phi) \bigr).     
\end{equation}     
Here $m$ and $n$ are co--prime integers, so the forcing function is     
periodic with period $T={2\pi\over \omega}$, and $g_0$ is the usual     
gravitational acceleration. For small amplitude acceleration $g_z$ the     
surface of the fluid remains flat and the fluid layer is merely     
translated up and down with the drive. For higher values of $g_z$     
waves are parametrically excited on the surface of the fluid layer.     
     
\begin{figure}      
\centerline{\resizebox{2.7in}{!} {\includegraphics{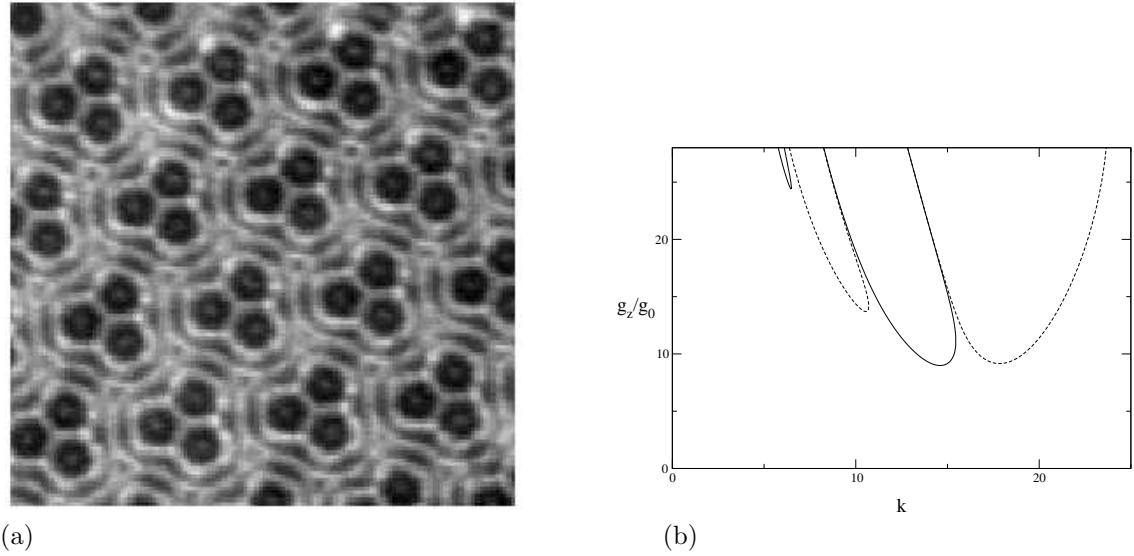}} \hspace{0.4in}    
\resizebox{2.7in}{!} {\includegraphics{67linearfhs.eps}}}    
(a) \hspace{3.2in} (b) 
\caption{(a) Blow up of the experimental superlattice Faraday    
wave pattern described in~\protect\cite{ref:kpg98} (courtesy of
Kudrolli, Pier and Gollub). The forcing
function~(\protect\ref{eq:f(t)}) has $m/n=6/7$, $\chi=61^\circ$ and
$\phi=20\degr$. Note that the pattern is periodic on a (large)
hexagonal lattice, and that in each hexagonal `tile' there is small
triangular structure. (b) The corresponding neutral stability curve,
calculated from the full (linearized) hydrodynamic equations, for the
experimental parameters reported in~\protect\cite{ref:kpg98}.
(Sub)harmonic resonance tongues are given by solid (dashed) lines.
The neutral curves are computed using the method described
in~\protect\cite{ref:bet96}.}
\label{fig:67sl}    
\end{figure}

Besson, Edwards and Tuckerman~\cite{ref:bet96}, starting with the 
Navier--Stokes equations for the free boundary problem, determined the 
linear stability of the flat surface in the case that the fluid layer 
has finite depth but is unbounded horizontally. They used a 
Floquet-Fourier ansatz and solved the linear stability problem 
numerically to determine, for each spatial wavenumber $k$, the value 
of $g_z$ where a Floquet multiplier first crosses the unit circle. The 
resulting neutral stability curves show that the primary instability 
is to either subharmonic or harmonic waves depending on the value of 
$\chi$ and the values of $m$ and $n$. (Harmonic/subharmonic response 
is relative to the forcing period $T=2\pi/\omega$.) Typically, if 
$\chi$ is small so that $\cos (\chi) \cos (m \omega t)$ is of greater 
significance than $\sin (\chi) \cos (n \omega t+\phi)$, then the 
response is harmonic if $m$ is even and subharmonic if $m$ is odd. 
Similarly, if $\chi$ is close to $\pi/2$, the primary instability is 
(sub)harmonic if $n$ is even (odd).  At the so--called bicritical 
point, $\chi=\chi_c$, both harmonic and subharmonic instabilities 
onset at the same value of the excitation amplitude, but with 
different wavenumbers. The harmonic superlattice pattern of 
Figure~\ref{fig:67sl}a, observed by Kudrolli, Pier and 
Gollub~\cite{ref:kpg98}, was obtained near the bicritical point for 
$m/n=6/7$ forcing in~(\ref{eq:f(t)}). The pertinent neutral stability 
curve, computed using the experimental fluid parameters, is given in 
Figure~\ref{fig:67sl}b. 
    
\subsection{Spatio--Temporally Resonant Triads}     
\label{sec:resonants}     
     
\begin{figure}     
\centerline{\resizebox{\textwidth}{!} {\includegraphics{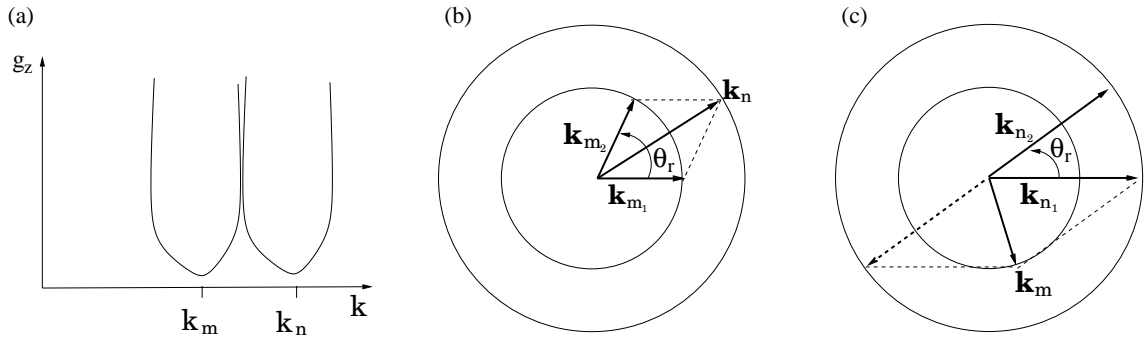}}}       
\caption{(a) A plot of a neutral stability  curve $g_z(k)$     
showing minima at $k=k_m$ and $k=k_n$. (b) An associated spatially 
resonant triad ${\bf k}_{m_1}$, ${\bf k}_{m_2}$ and ${\bf k}_n={\bf 
k}_{m_1}+{\bf k}_{m_2}$.  (c) An associated spatially resonant triad 
${\bf k}_{n_1}$, ${\bf k}_{n_2}$ and ${\bf k}_m={\bf k}_{n_1}-{\bf 
k}_{n_2}$.  } 
\label{fig:resonant-triad}     
\end{figure}

When the hydrodynamic problem is posed on a horizontally unbounded 
domain there is no preferred direction (in the horizontal) so that 
each critical wavenumber from linear analysis actually corresponds to 
a circle of critical wavevectors. There are two such critical circles 
at the bicritical point, as shown in Figure~\ref{fig:resonant-triad}. 
In this situation it has been argued that resonant triads may play a 
central role in the Faraday wave pattern selection 
problem~\cite{ref:ef94,ref:m93,ref:bwv97,ref:zv97b,ref:lp97}. Resonant 
triads are comprised of three critical wavevectors that sum to zero; 
two examples are shown in Figure~\ref{fig:resonant-triad}.  In the 
first example, ${\bf k}_{m_1} + {\bf k}_{m_2} - {\bf k}_n ={\bf 0}$, 
and in the second example ${\bf k}_{n_1} - {\bf k}_{n_2} - {\bf k}_m = 
{\bf 0}$.  Here the $m,n$ subscripts indicate that the critical 
wavenumbers can be roughly associated with the $m\omega$ and $n\omega$ 
excitation terms in~(\ref{eq:f(t)}). We identify with each resonant 
triad an angle $\theta_r\in (0,{\pi \over 2}]$, which separates the 
critical wavevectors with the same length. For instance, the angle in 
Figure~\ref{fig:resonant-triad}b satisfies 
\begin{equation}     
\label{eq:theta-r-b}     
\cos\Bigl({\theta_r\over 2}\Bigr)={k_n\over 2k_m}\ ,    
\end{equation}     
while the angle in Figure~\ref{fig:resonant-triad}c    
satisfies    
\begin{equation}     
\label{eq:theta-r-c}     
\sin\Bigl({\theta_r\over 2}\Bigr)={k_m\over 2k_n}\ .    
\end{equation}

To illustrate the potential for resonant triads to influence pattern 
formation in parametrically excited systems we consider a bifurcation 
problem involving the three critical Fourier modes associated with the 
resonant triads of Figure~\ref{fig:resonant-triad}. Much of this 
discussion is a review of the key theoretical ideas in~\cite{ref:ss99}. Because of the periodic 
forcing of the system, it is natural to formulate the bifurcation 
problem in terms of a stroboscopic map~\cite{ref:c90a}. Specifically, 
we denote the free surface height $z=h({\bf x},t)$ (${\bf x}\in{\bf 
R}^2$) at time $t=pT$ ($p\in{\bf Z}$) by 
\begin{equation}     
h({\bf x},pT)=A(p)e^{i{\bf k}_{l_1}\cdot{\bf x}}+B(p)e^{i{\bf     
k}_{l_2}\cdot{\bf x}}+C(p)e^{i({\bf k}_{l_1}+{\bf k}_{l_2})\cdot{\bf     
x}}+c.c.+\cdots \ .     
\end{equation}     
Here $A,B$ and $C$ are the complex amplitudes of the linear modes that     
are neutrally stable at the bicritical point and which form a resonant     
triad. In this discussion we assume that the angle $\theta_r$ between ${\bf     
k}_{l_1}$ and ${\bf k}_{l_2}$ is not $\pi/3$ so that the critical     
modes interact nonlinearly to generate other modes on a rhombic     
(rather than hexagonal) lattice.  These additional modes, denoted by     
$\cdots$ above, are linearly damped at the bicritical point.  We may     
then use the spatial reflection and translation symmetries to     
determine the general form of the bifurcation equations that govern     
the dynamics on a center manifold. Specifically, to cubic order, the     
codimension--two bifurcation problem takes the form     
\begin{eqnarray}     
\label{ABCinteraction}     
{A} &\to& \sigma A+\alpha     
\overline{B}C+(a|A|^2+b|B|^2+c|C|^2)A     
\nonumber\\     
{B} &\to& \sigma B+\alpha     
\overline{A}C+(a|B|^2+b|A|^2+c|C|^2)B\\     
{C} &\to& \mu C+\delta AB+(d|A|^2+d|B|^2+e|C|^2)C\ ,\nonumber     
\end{eqnarray}     
where $\overline{A}$ is the complex conjugate of $A$, and the     
coefficients are all real. The Floquet multipliers $\sigma$ and     
$\mu$ are either $+1$ or $-1$ depending on whether the linear modes     
$A$, $B$, and $C$ are harmonically or subharmonically excited, respectively.     
     
In deriving~(\ref{ABCinteraction}) we considered only the spatial
symmetries associated with the resonant
triad. Following~\cite{ref:c90a}, we enforce the temporal symmetry
associated with the triad through a normal form transformation
of~(\ref{ABCinteraction}).  Specifically, there exists a
near--identity nonlinear transformation that removes all nonlinear
terms in (\ref{ABCinteraction}) which do not commute with $L^T$, where
$L$ is the Jacobian matrix associated with the linearized problem
(see, for example, Crawford's review paper on bifurcation theory
\cite{ref:c91a}). Here
\begin{equation}     
\label{eq:nf-symmetry}     
L=\pmatrix{\sigma & 0 &0\cr 0 & \sigma &0 \cr 0 & 0 & \mu},     
\end{equation}     
where $|\sigma|=|\mu|=1$. The normal form symmetry may be interpreted     
in terms of time--translation. Specifically,  advancing     
by one period in time maps     
period--doubled modes to their negatives, {\it e.g.} if $\mu=-1$, then     
advancing one period takes $C\to -C$.     
     
In the case that $\mu=+1$ ($\sigma= \pm 1$), the bifurcation     
problem~(\ref{ABCinteraction}) is already in normal form. This     
observation is trivial if $\sigma=+1$. If $\sigma=-1$, then the     
normal form symmetry is equivalent in action to that associated with     
the spatial translation symmetry ${\bf x}\to{\bf x}+{\bf d}$, where     
${\bf d}$ satisfies ${\bf k}_{l_1}\cdot {\bf d}={\bf k}_{l_2}\cdot     
{\bf d}=\pi$.     
     
In contrast, in the case that $\mu=-1$, a normal form transformation 
removes the quadratic terms in the bifurcation 
problem~(\ref{ABCinteraction}). The normal form of the bifurcation 
problem, through cubic order, is then 
\begin{eqnarray}     
\label{ABC-pm-nf}     
{A} &\to& \sigma A+(a|A|^2+b|B|^2+c|C|^2)A     
\nonumber\\     
{B} &\to& \sigma B+(a|B|^2+b|A|^2+c|C|^2)B\\     
{C} &\to& - C+(d|A|^2+d|B|^2+e|C|^2)C\ .\nonumber     
\end{eqnarray}     
We note that $C=0$ is a dynamically--invariant subspace    
of~(\ref{ABC-pm-nf}). This is true to all orders of the normal form    
since $C=0$ is the fixed point subspace of a (spatio--)temporal    
symmetry. Specifically, if $\sigma=+1$ then $C=0$ is the fixed point    
subspace associated with the time translation by one--period, {\it    
i.e.}, $(A,B,C)\to(A,B,-C)$. And if $\sigma=-1$, then $C=0$ is the    
fixed--point subspace associated with the spatio--temporal symmetry    
involving time translation by one period followed by spatial    
translation by ${\bf d}$, where again ${\bf k}_{l_1}\cdot {\bf d}={\bf    
k}_{l_2}\cdot {\bf d}=\pi$.    
     
We now examine~(\ref{ABCinteraction}) more closely in the case that 
$\mu=+1$ so that we cannot remove the quadratic nonlinearities by 
normal form transformation. We focus on a detuning from the bicritical 
point such that the $C$ mode is weakly damped, while the $A,B$ modes 
are neutrally stable. In this case, $|\sigma|=1,\ \mu<1$, we can 
further reduce the bifurcation problem to one involving the critical 
modes $A$ and $B$, with $C$ constrained to the center manifold: 
$C={\delta\over (1-\mu)}AB+\cdots$. We then obtain the reduced 
bifurcation problem 
\begin{eqnarray}     
\label{eq:amps}     
{A} &\to& \sigma A+a|A|^2A     
      +\beta(\theta_r)|B|^2A\nonumber\\     
{B} &\to& \sigma B+a|B|^2B     
      +\beta(\theta_r)|A|^2B\ ,     
\end{eqnarray}     
where the cross--coupling coefficient is      
\begin{equation}     
\label{eq:cross-coupling}     
\beta(\theta_r)= b+{\alpha\delta\over(1-\mu)}\ .     
\end{equation}
We see that in this case, the near critical spatio--temporally
resonant mode $C$ in (\ref{ABCinteraction}) can contribute
significantly to the cross-coupling coefficient $\beta(\theta_r)$
since $0<1-\mu\ll 1$ in~(\ref{eq:cross-coupling}). For example, for
$\mu$ sufficiently close to 1, the second term
in~(\ref{eq:cross-coupling}) dominates and $\beta(\theta_r)$ becomes
large in magnitude. However, we also point out that if $b$ and
$\alpha\delta$ have opposite signs, then $\beta(\theta_r)$ could
actually vanish for some $\mu-1>0$. Examples of these two very
different situations are given in sections~\ref{sec:res23}
and~\ref{sec:res67}, respectively.

We contrast the above with what happens when $\mu=-1$ at the
bicritical point.  In this case $\alpha=\delta=0$ in the normal
form~(\ref{ABC-pm-nf}) and $C=0$ is an invariant subspace with
associated dynamics of the form~(\ref{eq:amps}) with
$\beta(\theta_r)=b$. In this case, the triad is spatially resonant,
but not temporally resonant, and the cross--coupling coefficient is
insensitive to any parameter proximity to the bicritical point.
 
These observations about $\beta(\theta_r)$ are important for
understanding which patterns might be observable near onset since
branching direction and stability of patterns are determined by
various nonlinear (cross--coupling) coefficients in the amplitude
equations.  We discuss this further at the end of
Section~\ref{sec:bifurcation}.
 
Finally we note that similar results to the $\mu=-1$ case above apply
when there are weakly damped modes with {\it complex} Floquet
multipliers. Specifically, these modes do not contribute significantly
to the cubic cross--coupling coefficient $\beta(\theta)$, even when
they are spatially resonant with the critical modes. Only damped modes
with Floquet multiplier $\mu$ sufficiently close to $+1$ contribute.
  
\subsection{Hexagonal Lattice Bifurcation Problem}      
\label{sec:bifurcation}      
     
The analysis of the previous section led to certain conclusions about     
the nonlinear coefficients in the general rhombic lattice bifurcation     
problem     
\begin{eqnarray}     
\label{eq:rhombs}     
v_1&\to&\sigma v_1 +(a |v_1|^2+\beta(\theta)|v_2|^2)v_1\nonumber\\     
v_2&\to&\sigma v_2 +(a |v_2|^2+\beta(\theta)|v_1|^2)v_2.    
\end{eqnarray}     
Here $v_1$, $v_2$ are the complex amplitudes of two critical Fourier 
modes with wavevectors ${\bf k}_1$, ${\bf k}_2$ ($|{\bf k}_1|=|{\bf 
k}_2|=k_c$) that are separated by an angle $\theta\in(0,{\pi\over 2}]$ 
($\theta\ne {\pi\over 3}$).  In particular, it follows 
from~(\ref{eq:cross-coupling}) that if a weakly damped harmonic mode 
is removed via center manifold reduction, then $\beta(\theta)$ may become 
large in magnitude when the spatial resonance condition is met, {\it 
i.e.} when $\theta=\theta_r$. This is in contrast to the situation 
where there are weakly damped subharmonic modes, which have no special 
influence on the pattern selection problem at onset. 
     
\begin{figure}[t]  
\vskip 0.8truein   
\setlength{\unitlength}{0.3000pt}
\begin{picture}(1007,500)(-150,0)
\thicklines
\put(530,456){\ellipse{438}{438}}
\multiput(175,167)(0,82.5){8}{\blacken\ellipse{15}{15}}
\multiput(246,126)(0,82.5){9}{\blacken\ellipse{15}{15}}
\multiput(317,167)(0,82.5){8}{\blacken\ellipse{15}{15}}
\multiput(388,126)(0,82.5){9}{\blacken\ellipse{15}{15}}
\multiput(459,167)(0,82.5){8}{\blacken\ellipse{15}{15}}
\multiput(530,126)(0,82.5){9}{\blacken\ellipse{15}{15}}
\multiput(600,167)(0,82.5){8}{\blacken\ellipse{15}{15}}
\multiput(671,126)(0,82.5){9}{\blacken\ellipse{15}{15}}
\multiput(742,167)(0,82.5){8}{\blacken\ellipse{15}{15}}
\multiput(813,126)(0,82.5){9}{\blacken\ellipse{15}{15}}
\multiput(884,167)(0,82.5){8}{\blacken\ellipse{15}{15}}
\path(530,456)(671,621)
\path(642,599)(664,613)(653,590)(642,599)
\path(530,456)(327,495)
\path(353,497)(327,495)(350,483)(353,497)
\path(530,456)(597,259)
\path(582,280)(597,259)(595,285)(582,280)
\path(530,456)(597,652)
\path(582,631)(597,652)(595,626)(582,631)
\path(530,456)(327,417)
\path(353,415)(327,417)(350,429)(353,415)
\path(530,456)(666,297)
\path(643,314)(666,297)(653,323)(643,314)
\path(530,456)(389,621)
\path(418,599)(396,613)(407,590)(418,599)
\path(530,456)(733,495)
\path(707,497)(733,495)(710,483)(707,497)
\path(530,456)(463,259)
\path(478,280)(463,259)(465,285)(478,280)
\path(530,456)(463,652)
\path(478,631)(463,652)(465,626)(478,631)
\path(530,456)(733,417)
\path(707,415)(733,417)(710,429)(707,415)
\path(530,456)(394,297)
\path(417,314)(394,297)(407,323)(417,314)
\put(680,615){${\bf K_{1}}$}
\put(255,520){${\bf K_{2}}$}
\put(580,200){${\bf K_{3}}$}
\put(250,390){${\bf K_{5}}$}
\put(590,675){${\bf K_{4}}$}
\put(690,275){${\bf K_{6}}$}
\put(565,540){\bf$\theta_h$}
\path(611.5,550.7)(601.7,558.4)(592.5,564.3)(582.8,569.3)(570.7,574.2)
\end{picture}  
\vskip -0.7truein   
\caption{Hexagonal ${\bf k}$--space lattice, with critical circle of radius 
$k_c$ superimposed. In this example $(n_1,n_2)=(3,2)$ 
in~(\protect\ref{eq:K's}), and the critical circle intersects twelve 
points that lie at the vertices of two hexagons rotated by $\theta_h$ 
relative to each other.} 
\label{fig:hexlatt}   
\end{figure}
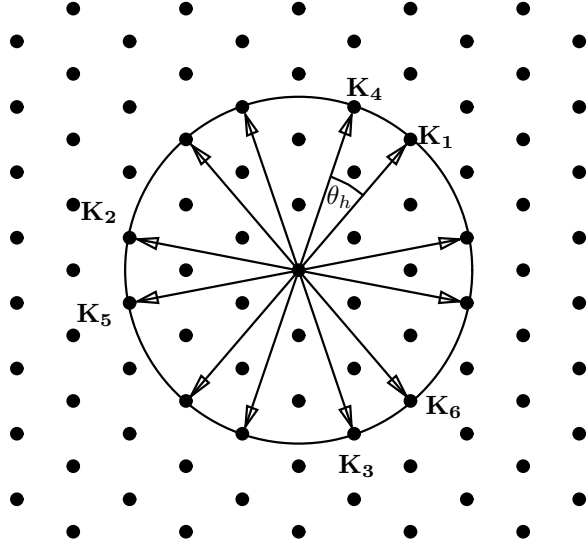

We now lay the framework for examining possible implications of these
results for stability of harmonic hexagonal and triangular
superpatterns. We follow~\cite{ref:sp98} and introduce the
twelve--dimensional ${\rm D}_6\dot{+}{\rm T}^2$--equivariant
bifurcation problems that enable us to determine the relative
stability of simple hexagonal patterns, stripe patterns and certain
rhombic and superlattice patterns. We make use of bifurcation results
derived in~\cite{ref:sp98,ref:dg92,ref:dss97}, which apply when there
is a {\it single} critical wavenumber $k_c$, to demonstrate how the
magnitude of the cross--coupling terms are pivotal in determining
pattern stability.  As before, we consider a stroboscopic map, but now
restrict analysis to patterns that are doubly--periodic on some
hexagonal lattice. For instance, the free surface height takes the
form
\begin{equation}     
\label{eq:fourier}     
h({\bf x},pT)=\sum_{{\bf m}\in{\bf Z}^2}     
\hat h_{{\bf m}}(p)e^{i(m_1{\bf k}_1+m_2{\bf k}_2)\cdot{\bf x}}+c.c.      
\end{equation}     
at time $t=pT$, where ${\bf k}_1, {\bf k}_2\in{\bf R}^2$ generate a 
hexagonal dual lattice ($|{\bf k}_1|=|{\bf k}_2|$ and ${\bf 
k}_1\cdot{\bf k}_2= -{1\over 2}|{\bf k}_1|^2$); see 
Figure~\ref{fig:hexlatt}. 
     
The twelve--dimensional irreducible representations of ${\rm 
D}_6\dot{+}{\rm T}^2$ apply to the bifurcation problem when there are 
twelve integer pairs $(m_1,m_2)$ in~(\ref{eq:fourier}) such that 
$|m_1{\bf k}_1+m_2{\bf k}_2|=k_c$, where $k_c$ is the critical 
wavenumber of the instability at the bifurcation point. See 
Figure~\ref{fig:hexlatt} for an example.  Following~\cite{ref:dg92} we 
will associate with each twelve--dimensional irreducible 
representation an integer pair $(n_1,n_2)$; in particular $n_1$ and 
$n_2$ are co--prime, $n_1>n_2>n_1/2>0$, and $n_1+n_2$ is not a 
multiple of 3. The neutral modes that span the center eigenspace at 
the bifurcation point take the form 
\begin{equation}     
\label{eq:critical-modes}     
\{z_1\ e^{i{\bf K}_1\cdot{\bf x}}+z_2\ e^{i{\bf K}_2\cdot{\bf x}}+     
z_3\ e^{i{\bf K}_3\cdot{\bf x}}+z_4\ e^{i{\bf K}_4\cdot{\bf x}}+     
z_5\ e^{i{\bf K}_5\cdot{\bf x}}+z_6\ e^{i{\bf K}_6\cdot{\bf x}}+c.c.|z_j\in     
{\bf C}\},     
\end{equation}     
where      
\begin{eqnarray}     
\label{eq:K's}     
{\bf K}_1=n_1{\bf k}_1+n_2{\bf k}_2,&\quad&      
{\bf K}_4=n_1{\bf k}_1+(n_1-n_2){\bf k}_2,\nonumber\\     
{\bf K}_2=(-n_1+n_2){\bf k}_1     
-n_1{\bf k}_2,&\quad&     
 {\bf K}_5=-n_2{\bf k}_1     
-n_1{\bf k}_2,\\     
{\bf K}_3=-n_2{\bf k_1}+(n_1-n_2){\bf k}_2,     
&\quad& {\bf K}_6=(n_2-n_1){\bf k_1}+n_2{\bf k}_2\ .\nonumber     
\end{eqnarray}     
Note that $\pm{\bf K}_1,\pm{\bf K}_2,\pm{\bf K}_3$ point to the 
vertices of a hexagon, as do $\pm{\bf K}_4,\pm{\bf K}_5,\pm{\bf K}_6$, 
and that the two hexagons are rotated relative to each other by an 
angle $\theta_h\in(0,{\pi\over 3})$ indicated in  
Figure~\ref{fig:hexlatt}. This angle is related to $(n_1,n_2)$ by  
\begin{equation}     
\label{eq:costheta}     
\cos(\theta_h)={n_1^2+2n_1n_2-2n_2^2\over 2(n_1^2-n_1n_2+n_2^2)}.      
\end{equation}     
Also note that the ratio of lengthscales for superpatterns depends 
on $(n_1,n_2)$. Specifically, $|{\bf k}_1|$ determines the larger 
periodicity scale of the superpatterns, while $|{\bf K}_j|=k_c$ 
determines the smaller lengthscale associated with the instability; 
thus the lengthscale ratio is 
\begin{equation}     
\label{eq:ratio}     
|{\bf K}_j|/|{\bf k}_1|=\sqrt{n_1^2-n_1n_2+n_2^2}\ge \sqrt{7}.      
\end{equation}      
The example of Figure~\ref{fig:hexlatt} corresponds to 
$(n_1,n_2)=(3,2)$, for which $\theta_h\approx 22\degr$ 
in~(\ref{eq:costheta}) and the lengthscale ratio (\ref{eq:ratio}) is 
the smallest associated with a hexagonal lattice, namely 
$\sqrt{7}$. These are the angle and lengthscale ratio that apply to 
the experimental superlattice pattern reproduced from~\cite{ref:kpg98} 
in Figure~\ref{fig:67sl}a. 
     
The general form of the twelve--dimensional ${\rm D}_6\dot{+}{\rm     
T}^2$--equivariant mappings are derived in~\cite{ref:dss97}. Through      
cubic order in $z_j$, they take the form     
\begin{eqnarray}     
\label{eq:bifurcation}     
{z}_1  \to  \sigma\Bigl(     
(1+\lambda) z_1 + \epsilon \overline{z}_2\overline{z}_3+     
(b_1|z_1|^2+b_2|z_2|^2+b_2|z_3|^2+b_4|z_4|^2+b_5|z_5|^2+b_6|z_6|^2)z_1     
\Bigr)&&\nonumber\\     
{z}_2  \to  \sigma\Bigl(     
(1+\lambda) z_2 + \epsilon \overline{z}_1     
\overline{z}_3+(b_1|z_2|^2+b_2|z_1|^2+b_2|z_3|^2+b_4|z_5|^2+b_5|z_6|^2     
+b_6|z_4|^2)z_2\Bigr)&&\nonumber\\     
{z}_3  \to  \sigma\Bigl(     
(1+\lambda) z_3 + \epsilon \overline{z}_1     
\overline{z}_2+(b_1|z_3|^2+b_2|z_1|^2+b_2|z_2|^2+b_4|z_6|^2+b_5|z_4|^2     
+b_6|z_5|^2)z_3\Bigr)&&\\     
{z}_4 \to  \sigma\Bigl(     
(1+\lambda) z_4 + \epsilon \overline{z}_5\overline{z}_6+     
(b_1|z_4|^2+b_2|z_5|^2+b_2|z_6|^2+b_4|z_1|^2+b_5|z_3|^2+b_6|z_2|^2)z_4     
\Bigr)&&\nonumber\\     
{z}_5 \to  \sigma\Bigl(     
(1+\lambda) z_5 + \epsilon \overline{z}_4\overline{z}_6+     
(b_1|z_5|^2+b_2|z_4|^2+b_2|z_6|^2+b_4|z_2|^2+b_5|z_1|^2+b_6|z_3|^2)z_5     
\Bigr)&&\nonumber\\     
{z}_6 \to  \sigma\Bigl(     
(1+\lambda) z_6 + \epsilon \overline{z}_4\overline{z}_5+     
(b_1|z_6|^2+b_2|z_4|^2+b_2|z_5|^2+b_4|z_3|^2+b_5|z_2|^2+b_6|z_1|^2)     
z_6\Bigr),&&\nonumber     
\end{eqnarray}     
where $\lambda$ measures the distance from the critical excitation    
amplitude, and $\sigma=+1$($-1$) in the case of (sub)harmonic    
instability. All nonlinear coefficients are real. If    
$\sigma=-1$ then a normal form transformation removes all even terms    
on the right--hand--side of (\ref{eq:bifurcation}) and hence 
$\epsilon=0$. The dependence of the general equivariant bifurcation    
problem on $(n_1,n_2)$ does not appear until higher than cubic order    
in its Taylor expansion~\cite{ref:dss97}.    
     
We now recall some basic results pertaining to the bifurcation 
problem~(\ref{eq:bifurcation}). In the $\sigma=+1$ case the 
equivariant branching lemma~\cite{ref:gss} ensures the existence of 
harmonic wave solution branches in the form of stripes, simple 
hexagons, rhombs, and super hexagons~\cite{ref:dg92}. A primary 
solution branch with submaximal isotropy, named super triangles, was 
also shown to exist in~\cite{ref:sp98}. See Figure~\ref{fig:67sl}a for 
an example of this pattern.  Table~\ref{tab:evhex} gives the general 
form of these solutions, along with their branching and stability 
assignments. The general bifurcation results in the case that 
$\sigma=-1$ can be found in~\cite{ref:dss97}; this bifurcation problem 
differs from the harmonic case in that it possesses an additional 
${\rm Z}_2$ normal form symmetry. The equivariant branching lemma then 
ensures existence of five additional solution branches to those listed 
in Table~\ref{tab:evhex}~\cite{ref:dss97}. 
     
\begin{centering}     
\begin{table}     
\caption{Branching equations and stability assignments for the 
harmonic case ($\sigma=+1$); $\epsilon, b_1,\dots,b_6$ are 
coefficients in the bifurcation 
equations~(\protect\ref{eq:bifurcation}). A solution is stable if all 
quantities in the right column are negative. See 
\protect\cite{ref:sp98,ref:dg92,ref:dss97} for more details. 
} 
\label{tab:evhex}      
\begin{tabular}{|c|c|}      
\hline      
Planform and branching equation	& Stability \\       
\hline      
& \\       
Stripes      
       & $\sgn(b_1)$,      
\\       
${\bf z}=(x,0,0,0,0,0)$ &      
 	$\sgn(\epsilon x+(b_2-b_1)x^2), \quad \sgn(-\epsilon x+(b_2-b_1)x^2)$,\\      
$0=\lambda x +b_1 x^3+{\mathcal O}(x^5)$ &                  
$\sgn(b_4 - b_1), \quad \sgn(b_5 - b_1), \quad \sgn(b_6 - b_1)$\\      
& \\      
\hline      
& \\       
Simple Hexagons      
       & $ \sgn(\epsilon x+2(b_1+2b_2)x^2), \quad       
\sgn(-\epsilon x+(b_1-b_2)x^2$)               \\       
${\bf z}=(x,x,x,0,0,0)$ &      
 $ \sgn(-\epsilon x+(b_4+b_5+b_6-b_1-2b_2)x^2)$  \\       
$0=\lambda x +\epsilon x^2 + (b_1 + 2b_2) x^3 + {\mathcal O}(x^4)$      
       & $ \sgn(-\epsilon x+{\mathcal O}(x^3))$          \\       
 &\\      
\hline      
& \\        
Rhombs (Rh$_4$)      
       & $\sgn(b_1 + b_4), \quad \sgn(b_1 - b_4),\quad \sgn(\zeta_1),      
		\quad \sgn(\zeta_2),$ \\       
${\bf z}=(x,0,0,x,0,0)$ &      
	where $\zeta_1+\zeta_2=(-2b_1-2b_4+2b_2+b_5+b_6)x^2$, \\       
$0=\lambda x + (b_1+b_4) x^3+{\mathcal O}(x^5)$      
	& $\zeta_1\zeta_2=      
	  -\epsilon^2x^2+(b_1+b_4-b_2-b_5)(b_1+b_4-b_2-b_6)x^4$ \\       
& \\       
\hline      
& \\       
Rhombs (Rh$_5$)      
       &  same as Rh$_4$ with $b_4\leftrightarrow b_5$ \\       
${\bf z}=(x,0,0,0,x,0)$      
	&      
 \\        
& \\       
\hline      
& \\       
Rhombs (Rh$_6$)      
       &  same as Rh$_4$ with $b_4\leftrightarrow b_6$ \\       
${\bf z}=(x,0,0,0,0,x)$      
	&      
 \\        
& \\       
\hline     
&\\      
&$\sgn(\epsilon x+2(b_1+2b_2+b_4+b_5+b_6)x^2)$\\     
Super Hexagons      
	& $\sgn(\epsilon x+2(b_1+2b_2-b_4-b_5-b_6)x^2)$ \\       
${\bf z}=(x,x,x,x,x,x)$       
	& $\sgn(-\epsilon x+{\mathcal O}(x^3)),      
	  \quad \sgn(\zeta_1),\quad \sgn(\zeta_2)$,\\      
$0=\lambda x+\epsilon x^2+(b_1+2b_2)x^3\quad$       
	& where $\zeta_1+\zeta_2=-4\epsilon x +4(b_1-b_2)x^2$, \\       
$\qquad+(b_4+b_5+b_6)x^3+{\mathcal O}(x^4)$      
	& $\zeta_1\zeta_2=4(\epsilon x-(b_1-b_2)x^2)^2\qquad\qquad\qquad\quad$      
\\      
&$\qquad \qquad\qquad-2((b_4-b_5)^2+(b_4-b_6)^2+(b_5-b_6)^2))x^4$\\      
&$\sgn(\zeta_3)$, where       
$\zeta_3={\mathcal O}(x^{2(n_1-1)})$\\ & \\      
\hline      
& \\      
Super Triangles &  Same as super hexagons \\      
${\bf z} = (z,z,z,z,z,z),$ & except $\zeta_3 \to -\zeta_3$\\      
 $\quad z=x e^{i\psi}, \psi\ne 0,\pi,\ldots$ & \\      
& \\       
\hline      
\end{tabular}      
\end{table}       
 \end{centering}     
     
The generic presence of a quadratic term in~(\ref{eq:bifurcation}) for
the harmonic case renders all of the solutions in
Table~\ref{tab:evhex} unstable at bifurcation. Hence the transition
from the flat state to the patterned harmonic wave state is expected
to be hysteretic. In order to capture stable weakly nonlinear
solutions, we must focus our analysis on the unfolding of the
degenerate bifurcation problem $\epsilon=0$. Note that when
$\epsilon=0$ the stability of simple and super hexagons/triangles is
not determined at cubic order since the phases $\phi_j$ of solutions
$z_j=r_je^{i\phi_j}$ to~(\ref{eq:bifurcation}) are then arbitrary.
Even in the case of $0<|\epsilon|\ll 1$ the relative stability of
super hexagons and super triangles depends on terms that are at least
fifth order. However, we may use the cubic truncation to determine
that one (and only one) of these two solutions is stable. The higher
order terms are only needed to determine whether it is the hexagonal
or triangular superpattern~\cite{ref:sp98}.
     
When $0<|\epsilon|\ll 1$, it follows from Table~\ref{tab:evhex} that a    
necessary condition for one of the superpatterns to be stable over    
some range of $\lambda$ values near onset is for    
\begin{equation}     
\label{eq:conditions}     
b_1+2b_2 < -|b_4+b_5+b_6|<0 \ .     
\end{equation}     
The combination $b_1+2b_2$ is independent of the lattice angle    
$\theta_h$ in~(\ref{eq:costheta}); it is computed from a hydrodynamic    
model of the two--frequency Faraday problem in the Appendix by    
considering bifurcation to simple hexagons.  In contrast, the    
combination $b_4+b_5+b_6$ depends on $\theta_h$ and is computed in the    
appendix from the hydrodynamic equations by considering the rhombic    
lattice bifurcation problem~(\ref{eq:rhombs}). Specifically, the    
cross--coupling coefficients $b_4,b_5,b_6$ are    
\begin{equation}     
\label{eq:b's}     
b_4=\beta(\theta_h),\quad b_5=\beta\Bigl(\theta_h+{2\pi\over 3}\Bigr),     
\quad b_6=\beta\Bigl(\theta_h-{2\pi\over 3}\Bigr),     
\end{equation}     
where $\theta_h$ is the angle between ${\bf K}_1$ and ${\bf K_4}$    
given by~(\ref{eq:costheta}). (The function $\beta(\theta)$ may be    
extended from $\theta\in(0,{\pi\over 2}]$ to angles    
$\theta\in(0,2\pi)$ using $\beta(\theta)=\beta(-\theta)=\beta(\theta    
+\pi)$, identities that follow from the symmetries of the rhombic    
lattice bifurcation problem.)    
    
The inequality~(\ref{eq:conditions}) will be satisfied (if at all)    
only for those $\theta_h$ values where $|b_4+b_5+b_6|$ is small    
compared to $|b_1+2b_2|$.  Moreover, if $b_1-b_2<0$ in addition    
to~(\ref{eq:conditions}), then simple hexagons become unstable on a    
given hexagonal lattice when    
\begin{equation}     
\label{eq:simp-lam}     
\lambda=     
-{\epsilon^2(b_4+b_5+b_6)\over (b_1+2b_2-b_4-b_5-b_6)^2}\ .     
\end{equation}     
If $b_4+b_5+b_6<0$ for all $\theta_h$, then simple hexagons first lose 
stability with increasing $\lambda$ to a perturbation in the direction 
of a superpattern for that value of $\theta_h$ that minimizes 
$|b_4+b_5+b_6|$. If $b_4+b_5+b_6>0$ for any $\theta_h$, then small 
amplitude simple hexagons are unstable when $\lambda>0$.  Thus we 
expect the stability properties of superpatterns and simple hexagons 
to be affected by the presence of a weakly damped harmonic mode when 
$\theta_h$ or $\theta_h\pm 2\pi/3$ is near $\theta_r$ (or 
$\pi\pm\theta_r$), the resonant triad angle, since it is in this 
situation that one of the cross--coupling coefficients $b_4$, $b_5$ or 
$b_6$ may suddenly change in magnitude.


\section{Results}      
\label{sec:results}      
      
      
      
      
      
\begin{figure}
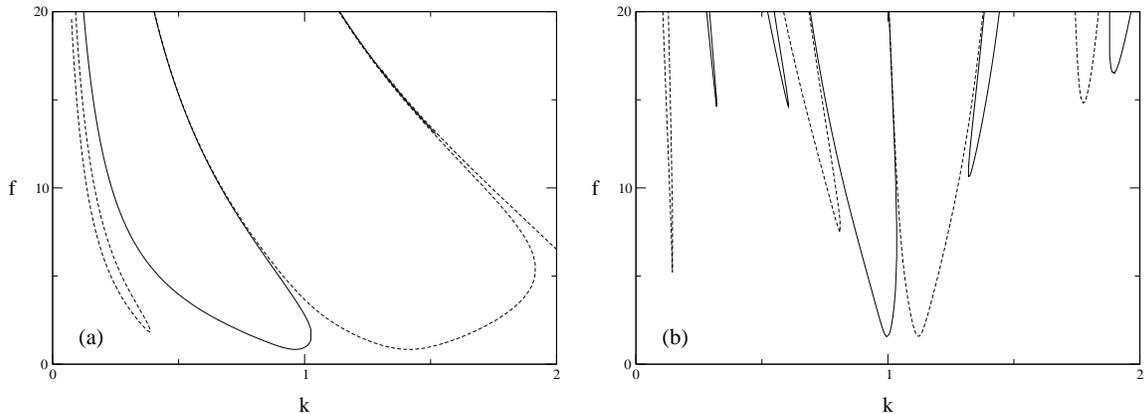
      
\centerline{\resizebox{\textwidth}{!} {\includegraphics{23linear.eps}    
\hspace{0.4in} \includegraphics{67hlinear.eps}}}    
\caption{Neutral stability curves computed     
from~(\protect\ref{eq:ZVmodel}) linearized about $h=\Phi=0$.  Floquet 
multipliers of $+1$ ($-1$) are indicated by solid (dashed) lines.  (a) 
$m/n=2/3$, $\phi=0\degr$, $\chi=\chi_c=66.6\degr$, $\Gamma_0=0.53$, 
$G_0=0.47$ and $\gamma=0.09$ 
in~(\protect\ref{eq:ZVmodel})--(\protect\ref{eq:G(t)}). (b) $m/n=6/7$, 
$\phi=0\degr$, $\chi=\chi_c=53.0\degr$, $\Gamma_0=7.5$, $G_0=1.5$ and 
$\gamma=0.08$.  } 
\label{fig:linear}     
\end{figure}

This section shows explicitly the role of resonant triads and weakly 
damped harmonic modes in the pattern selection problem for 
two--frequency forced Faraday waves. We examine how the cubic nonlinear 
coefficients in~(\ref{eq:bifurcation}),  
for the Zhang--Vi\~nals hydrodynamic equations vary as a function 
of $\theta_h$, the lattice angle and explain how this 
can be related to $\theta_r$, the resonant triad angle. 
The details of the computation of the coefficients 
are relegated to the Appendix. 
We focus on two examples, involving forcing frequency 
ratios $m/n=2/3$ and 6/7. The 2/3 case demonstrates the basic 
difference between the pattern selection problems for subharmonic 
and harmonic instabilities near the bicritical point. Our investigation 
also reveals a fundamental difference between harmonic wave pattern 
selection in the 2/3 and 6/7 cases, due to the presence of additional 
harmonic wave resonance tongues for the higher 6/7 forcing 
frequencies; see Figure~\ref{fig:linear}. 
     
\subsection{The Zhang-Vi\~{n}als Hydrodynamic Equations}     
\label{sec:zveq}      
     
The quadratic and cubic nonlinear coefficients in the hexagonal 
bifurcation problem~(\ref{eq:bifurcation}) are computed in the 
appendix from a model of the two--frequency Faraday problem derived by 
Zhang and Vi\~nals~\cite{ref:zv97a} from the Navier--Stokes equations.
Their equations, which apply to weakly damped, small amplitude 
surface waves on a semi--infinite layer of fluid, describe the 
evolution of the surface height $h({\bf x},\tau)$ and surface velocity 
potential $\Phi({\bf x},\tau)$. Specifically, 
\begin{eqnarray}     
\label{eq:ZVmodel}     
\partial_\tau h &=&      
\gamma\nabla^2h + \widehat \mathcal{D} \Phi      
-\nabla \cdot( h \nabla     
\Phi) + {1\over 2} \nabla^2 (h^2 \widehat \mathcal{D} \Phi)     
- \widehat \mathcal{D} (h \widehat \mathcal{D} \Phi)     
+\widehat \mathcal{D} [h\widehat \mathcal{D}(h\widehat     
\mathcal{D}\Phi)+{1\over 2}h^2\nabla^2{\Phi}] \nonumber \\     
\partial_\tau\Phi&=&\gamma\nabla^2 \Phi + \Gamma_0\nabla^2 h -G(\tau)h +      
{1\over 2}(\widehat     
D\Phi)^2-{1\over 2}(\nabla \Phi)^2-(\widehat \mathcal{D}     
\Phi)[h\nabla^2\Phi+\widehat \mathcal{D}(h\widehat \mathcal{D}\Phi)]     
\\	     
&&\mbox{} -{1\over 2}\Gamma_0\nabla\cdot( (\nabla h)(\nabla h)^2 ),     
\nonumber     
\end{eqnarray}     
where $\widehat \mathcal{D}$ is a nonlocal operator that multiplies     
each Fourier component of a field by its wave number, {\it i.e.}     
$\widehat \mathcal{D} e^{i{\bf k}\cdot{\bf x}}= |{\bf k}|e^{i{\bf     
k}\cdot{\bf x}}$. Here time has been scaled by $\omega$ so that the     
(non--dimensionalized) two--frequency acceleration is     
\begin{equation}     
\label{eq:G(t)}     
G(\tau)=G_0-f(\cos(\chi)\cos(m\tau)+\sin(\chi)\cos(n\tau+\phi)).     
\end{equation}     
The damping number ($\gamma$), capillarity number ($\Gamma_0$),     
gravity number ($G_0$), and dimensionless acceleration ($f$) are     
related to the forcing function~(\ref{eq:f(t)}) and the fluid     
parameters by     
\begin{equation}     
\gamma\equiv {2\nu k_0^2\over \omega}, \quad \Gamma_0 \equiv {\Gamma k_0^3\over     
\rho\omega^2}, \quad G_0\equiv{g_0 k_0 \over \omega^2}, \quad f\equiv {g_z     
k_0\over \omega^2}.     
\end{equation}     
Here $\nu$ is the kinematic viscosity, $\Gamma$ is the     
surface tension, $\rho$ is the fluid density, and the wave number     
$k_0$ is chosen to satisfy the dispersion relation     
\begin{equation}     
g_0 k_0+{\Gamma k_0^3\over \rho}=\Bigl({m\omega\over 2}\Bigr)^2.     
\end{equation}     
     
\subsection{Example 1: m/n=2/3}     
\label{sec:res23} 
     
This example demonstrates a result of the general normal form analysis    
of Section~\ref{sec:resonants}, namely that proximity to the    
subharmonic/harmonic bicritical point will strongly influence the    
pattern selection problem for subharmonic waves, but not for harmonic    
waves. Specifically, we examine the cross--coupling coefficient    
$\beta(\theta)$ in~(\ref{eq:rhombs}) as a function of the angle    
$\theta$ for onset of both harmonic and subharmonic waves near the    
bicritical point. We show that only in the subharmonic case does    
$|\beta(\theta)|$ become large at the resonant angle $\theta_r$    
in~(\ref{eq:theta-r-c}).    
     
As described in Section~\ref{sec:linear}, the primary instability 
changes from harmonic (Floquet multiplier $+1$) to subharmonic 
(Floquet multiplier $-1$) as $\chi$ in~(\ref{eq:G(t)}) is increased 
through the bicritical point $\chi_c$. This transition is determined 
from the linear hydrodynamic problem, which for the Zhang--Vi\~nals 
model~(\ref{eq:ZVmodel}) takes the form of a damped Mathieu equation 
for each Fourier mode $h=h_k(\tau)e^{ikx}$: 
\begin{equation}     
\label{eq:forced-linear}  h_k'' + 2\gamma k^2 h_k' + (\gamma^2    
k^4+\Omega_k^2)h_k=f_c k\bigl[    
\cos(\chi)\cos(m\tau)+\sin(\chi)\cos(n\tau)\bigr]h_k.    
\end{equation}    
Here the natural frequency $\Omega_k$ satisfies the dispersion 
relation $\Omega_k^2=G_0k+\Gamma_0k^3$.  A numerically--computed 
neutral curve $f(k)$ for $m/n=2/3$ forcing and $\chi=\chi_c=66.6\degr$ 
is given in Figure~\ref{fig:linear}a. The other parameters of this 
example are $\phi=0\degr$, $\Gamma_0=0.53$, $G_0=0.47$ and $\gamma=0.09$. 
     
We now vary $\chi$ near $\chi_c$, holding all other    
parameters fixed, and examine the rhombic lattice cross--coupling    
coefficient $\beta(\theta)$ in~(\ref{eq:rhombs}) for onset    
subharmonic/harmonic waves, as appropriate. We have scaled the    
amplitudes $v_1$ and $v_2$ in~(\ref{eq:rhombs}) so that $a=-1$. We    
note that in the harmonic case $\beta$ diverges as $\theta\to    
60\degr$, {\it i.e.} when the rhombic lattice approaches the hexagonal    
one and there is an additional mode associated with the center    
manifold dynamics. This is in contrast to the subharmonic case, for    
which there is a normal form symmetry that ensures existence of a    
dynamically invariant subspace spanned by a pair of subharmonic modes    
separated by $60\degr$. Thus in the subharmonic case $\beta$ remains    
finite at $\theta=60\degr$.    
    
\begin{figure}
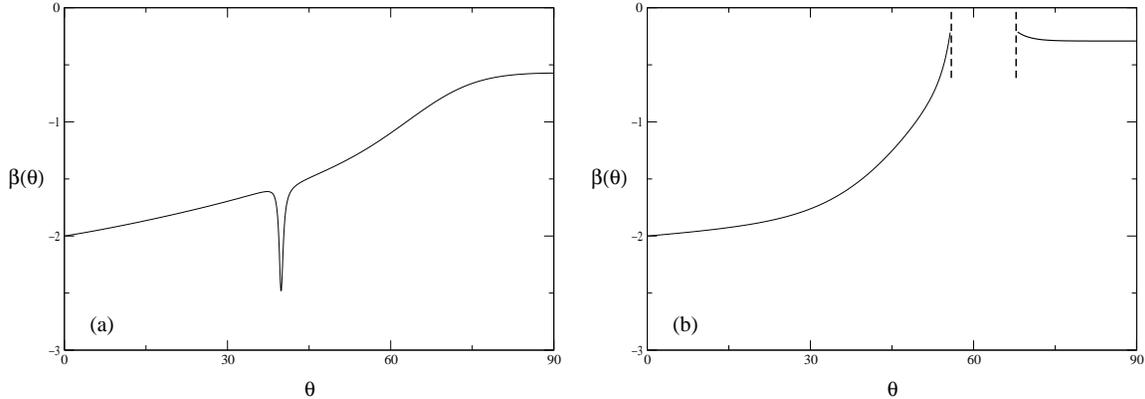
      
\centerline{\resizebox{\textwidth}{!}     
{\includegraphics{23shcross.eps} \hspace{0.4in}    
\includegraphics{23hcross.eps}}}       
\caption{Cross coupling coefficients  
$\beta(\theta)$ in~\protect(\ref{eq:rhombs}) computed in the Appendix 
from~(\protect\ref{eq:ZVmodel}) for the case $m/n=2/3$ and $\phi=0\degr$ 
in~(\protect\ref{eq:G(t)}).  The fluid parameters used are given in 
the caption of Figure~\protect\ref{fig:linear}a.  (a) 
$\chi=66.7\degr>\chi_c$, when the bifurcation is to subharmonic waves. 
Note the dip at $\theta=\theta_r=39.9\degr$.  (b) 
$\chi=66.5\degr<\chi_c$, when the bifurcation is to harmonic waves. 
Because the (nearly) critical modes are not in temporal resonance, 
$\beta(\theta)$ shows no special structure at 
$\theta=\theta_r=85.7\degr$. We have removed from this plot the region 
near $\theta=60\degr$, where $\beta(\theta)$ diverges.} 
\label{fig:23figs}     
\end{figure}

For $\chi>\chi_c$ the primary instability is to subharmonic waves. 
For instance, for $\chi= 66.7\degr$ the minimum of the neutral curve 
occurs at wavenumber $k_{c,s}=1.415$ with forcing amplitude 
$f_{c}=0.842$, and is associated with a Floquet multiplier $\sigma= 
-1$.  The nearly critical harmonic resonance tongue has its minimum at 
$(k,f)=(0.962,0.846)$. In this case, there is a spatio--temporally 
resonant triad comprised of the weakly damped harmonic mode and, 
from~(\ref{eq:theta-r-c}), two subharmonic modes separated by 
$\theta_r=39.9\degr$. It follows from our general analysis of 
Section~\ref{sec:resonants} that $\beta(\theta)$ will be large in 
magnitude for $\theta$ near $\theta_r$.  Figure~\ref{fig:23figs}a 
shows $\beta(\theta)$ for this case, and indeed, the nonlinear 
coefficient exhibits a large dip centered at 
$\theta=\theta_r=39.9\degr$. At this angle, $|\beta(\theta)|$ takes on 
its largest value. Similar observations have been made by Zhang and 
Vi\~{n}als~\cite{ref:zv97b} for forcing frequencies in ratio 
$m/n=1/2$.

In contrast, when $\chi<\chi_c$, so that the first instability to 
occur with increasing $f$ is harmonic, we find that the weakly damped 
subharmonic mode leaves no signature in the plot $\beta(\theta)$. For 
instance, for $\chi= 66.5\degr$ the primary instability is to harmonic 
waves at wavenumber $k_{c,h}=0.963$ and forcing amplitude 
$f_{c}=0.841$.  The subhharmonic resonance tongue has a minimum at 
$(k,f)=(1.415,0.843)$.  While there is a {\it spatially} resonant 
triad involving two critical harmonic modes, which 
by~(\ref{eq:theta-r-b}) are separated by $\theta_r=85.7\degr$, the 
triad of modes is not {\it spatio--temporally} 
resonant. Figure~\ref{fig:23figs}b shows the cross-coupling 
coefficient $\beta(\theta)$ for this case (with the region near 
$60\degr$ removed). As anticipated, there is no signature of the 
weakly damped subharmonic mode in the plot. Similar observations have 
been made by Silber and Skeldon~\cite{ref:ss99} in the setting of 
one--dimensional surface wave patterns. 
    
\subsection{Example 2: m/n=6/7}       
\label{sec:res67} 
    
This example demonstrates a fundamental difference between harmonic
wave pattern selection for low forcing frequencies ({\it e.g.},
$2\omega/3\omega$) and for high forcing frequencies ({\it e.g.},
$6\omega/7\omega$). This difference is due to the presence of multiple
harmonic resonance tongues in the neutral curve associated with the
higher forcing frequencies; see Figure~\ref{fig:linear}. In
particular, these resonance tongues suggest the possibility that weakly
damped harmonic modes may influence the {\it harmonic} wave pattern
selection problem. This is in contrast to the $m/n=2/3$ example of the
previous section, for which only {\it subharmonic} wave pattern
competition was affected by weakly damped harmonic waves. In this
section we also demonstrate that the weakly damped harmonic modes may
stabilize harmonic wave superpatterns at a lattice angle
$\theta_h\approx\theta_r$, due to a near cancellation of the two terms
that contribute to $\beta(\theta_r)$ given by~(\ref{eq:cross-coupling})
as described in Section~\ref{sec:resonants}.
    
    
We focus on bifurcation to harmonic waves for $\chi=52.4\degr$, which
is close to the bicritical value $\chi_c=53.0\degr$. The remaining
parameters are $\phi=0\degr$, $\Gamma_0=7.5$, $G_0=1.5$ and
$\gamma=0.08$.  We note that while the forcing frequency ratio
$m/n=6/7$ coincides with that used in the experiments of Kudrolli,
{\it et al.}~\cite{ref:kpg98}, the remaining parameters do not
coincide with the experiment. One problem with using the experimental
parameters in the Zhang--Vi\~{n}als equations is that the primary
instability then moves to a subharmonic resonance tongue at very small
wavenumber, {\it i.e.}, the first resonance tongue of
Figure~\ref{fig:linear}b. This is because the Zhang--Vi\~{n}als model
does not accurately capture the damping at small $k$ that is due to
finite depth effects.

\begin{figure}
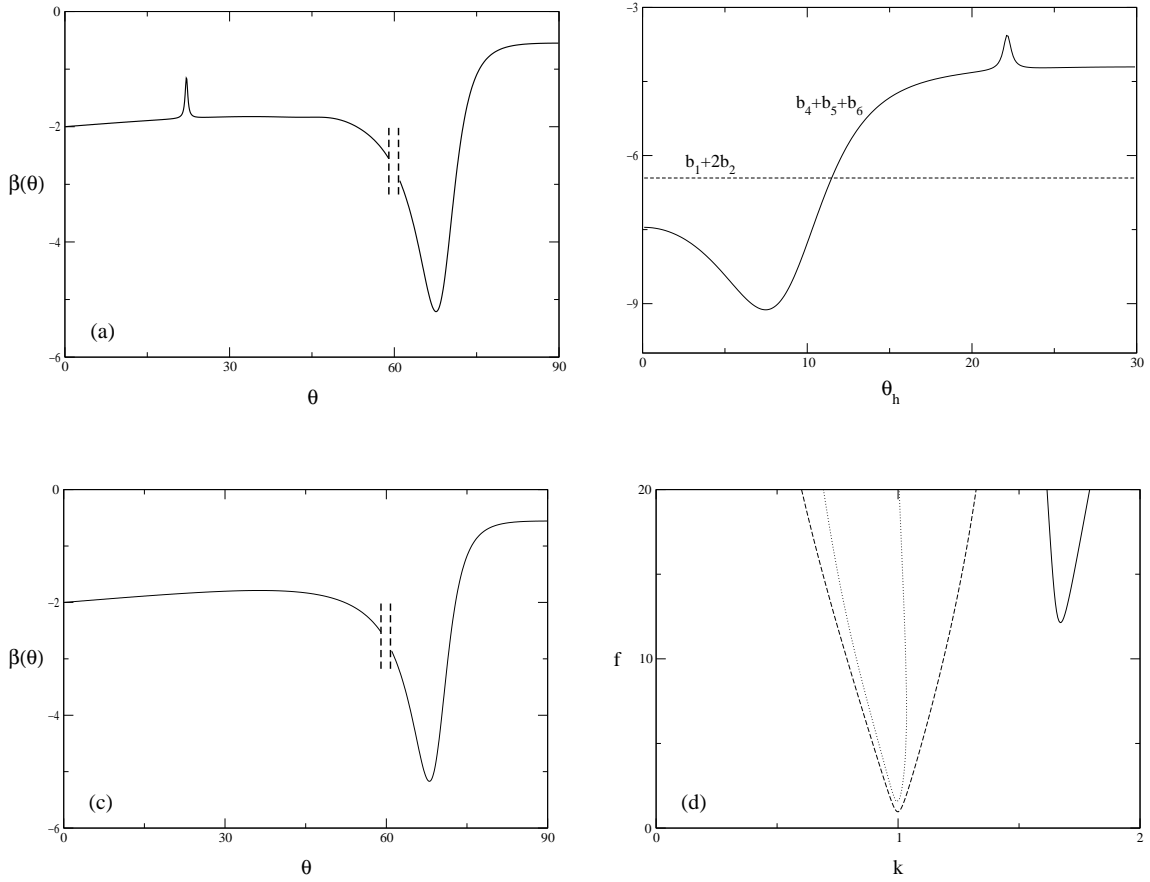
      
\centerline{\resizebox{\textwidth}{!} {\includegraphics{67hcross.eps}  
\hspace{0.9in} \includegraphics{b4b5b6.eps}}}    
\vspace{0.4in}    
\centerline{\resizebox{\textwidth}{!} {\includegraphics{6cross.eps}  
\hspace{0.9in} \includegraphics{6linear.eps}}}    
\caption{  
(a) Cross-coupling coefficient $\beta(\theta)$ in~(\protect 
\ref{eq:rhombs}) computed from~(\protect\ref{eq:ZVmodel}) for the 
case $m/n=6/7$, $\phi=0\degr$ and $\chi=52.4\degr<\chi_c$ 
in~(\protect\ref{eq:G(t)}), and for fluid parameters given in the 
caption of Figure~\protect\ref{fig:linear}b.  (b) Plots of $b_1+2b_2$ 
(dashed) and $b_4+b_5+b_6$ (solid) versus $\theta_h$. We note that 
$\theta_h$ only takes on the discrete values 
satisfying~(\protect\ref{eq:costheta}). (c) Cross-coupling coefficient 
$\beta(\theta)$ for $6\omega$ forcing only; we have used the same 
parameters as in (a) except that now $\chi=0\degr$.  (d) Neutral curve 
for single frequency forcing. Floquet multipliers of $+1$ ($-1$) are 
indicated by solid (dashed) lines, and are computed relative to the 
period $T'=2\pi/6$.  The primary harmonic resonance tongue from the 
two--frequency case of Figure~\protect\ref{fig:linear}b is 
superimposed as a dotted line.} 
\label{fig:67nonlinear}     
\end{figure}

In this example we find two prominent features in the plot of the
cross--coupling coefficient $\beta(\theta)$ in
Figure~\ref{fig:67nonlinear}a: a large dip at $\theta=67.6\degr$ and a
small spike at $\theta=22.2\degr$.  We now discuss the origin of these
two features.
    
The large dip around $\theta=67.6\degr$ is not a consequence of 
two--frequency forcing. Specifically, the dip remains in 
$\beta(\theta)$ even for purely $6\omega$ forcing ({\it i.e.}  in the 
limit $\chi\to 0$); {\it cf.} plots of $\beta(\theta)$ in 
Figures~\ref{fig:67nonlinear}a and~\ref{fig:67nonlinear}c which are 
obtained with $\chi=52.4\degr$ and $\chi=0\degr$, respectively. Thus 
this feature may be understood in the context of single frequency 
forcing, and has in fact already been investigated by Zhang and 
Vi\~{n}als~\cite{ref:zv97a} in that setting. Specifically, if 
$\chi=0\degr$ then the forcing period is $T'={T\over 6}={2\pi\over 6}$ 
and the primary instability is to subharmonic waves with period 
$2T'$. A plot of the corresponding neutral curve is given in 
Figure~\ref{fig:67nonlinear}d, with the primary harmonic resonance 
tongue from Figure~\ref{fig:linear}b superimposed on it. In this 
single--frequency setting the feature at $67.6\degr$ is understood as 
being due to the damped harmonic mode around $k=1.7$ in 
Figure~\ref{fig:67nonlinear}d. Perhaps more relevant 
to this discussion is our observation that this feature, which leads 
to a large value of $|b_4+b_5+b_6|$, is {\it destabilizing} for 
superpatterns. To see this, we refer to the discussion surrounding 
equation~(\ref{eq:conditions}) and to Figure~\ref{fig:67nonlinear}b, 
which shows that 
\begin{eqnarray}   
\label{eq:67.6bs}   
0>b_1+2b_2>b_4+b_5+b_6&=&\beta(\theta_h)+\beta\Bigl(\theta_h+{2\pi\over 
3}\Bigr)+\beta\Bigl(\theta_h-{2\pi\over 3}\Bigr),\nonumber\\ 
&=&\beta(\theta_h)+\beta\Bigl({\pi\over 
3}-\theta_h\Bigr)+\beta\Bigl({\pi\over 3}+\theta_h\Bigr) 
\quad {\rm for} \quad 
{\pi\over 3}+\theta_h\approx 67.6\degr. 
\end{eqnarray}

In contrast the spike at $\theta=22.2\degr$ in 
Figure~\ref{fig:67nonlinear}a minimizes $|b_4+b_5+b_6|$ at 
$\theta_h\approx 22.2\degr$, as shown in 
Figure~\ref{fig:67nonlinear}b. As we show below, this feature can lead 
to a stabilization of superpatterns and a destabilization of the 
simple hexagons. First we provide strong evidence that the spike is 
due to a resonance between the primary harmonic instability 
($\sigma=1$) and a weakly damped harmonic mode with a real Floquet 
multiplier $\mu$ that is close to 1 (see~(\ref{ABCinteraction}) 
and~(\ref{eq:cross-coupling}) of Section~\ref{sec:resonants}).  In 
order to show this we must first compute the Floquet multipliers 
$\mu(k)$ at the critical forcing amplitude $f_c$ to determine the 
wavenumbers $k$ at which $\mu\approx 1$ at the onset of instability. 
    
    
\begin{figure}
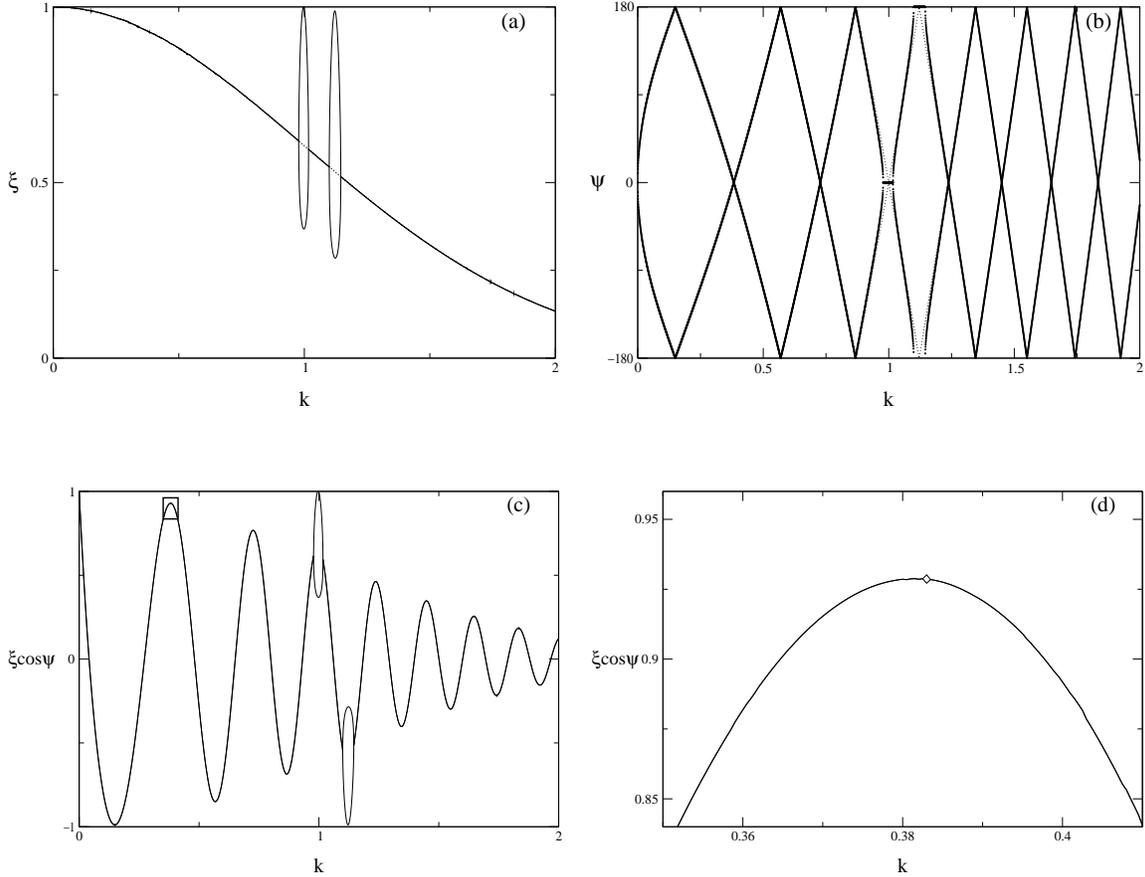
      
\centerline{\resizebox{\textwidth}{!} {\includegraphics{67hfmabs.eps}    
\hspace{0.4in} \includegraphics{67hfmphase.eps}}}    
\vspace{0.4in}    
\centerline{\resizebox{\textwidth}{!} {\includegraphics{67hfmreal.eps}    
\hspace{0.4in} \includegraphics{67hfmrealblowup.eps}}}    
\caption{Floquet multipliers $\mu=\xi e^{i\psi}$ computed  
from~(\protect\ref{eq:forced-linear}) for the parameters used in 
Figure~\protect\ref{fig:linear}b and for critical forcing amplitude 
$f=f_c=1.552$. (a) Magnitude $\xi$, and (b) phase $\psi$ {\it vs.} 
wavenumber $k$. The solid lines in (a) and (b) are computed 
numerically, while the the dotted lines are obtained by considering 
the unforced problem $f=0$; see equation~\protect 
\ref{eq:approxfm}.  (c) Numerically computed real part $\xi\cos \psi$ 
of the Floquet multipliers.  The ``bubbles" correspond to real-valued 
Floquet multipliers.  The boxed region, shown blown up in (d), reveals 
a tiny ``bubble" around $k=0.383$, with real Floquet multiplier 
$\mu=0.93$.} 
\label{fig:67hfm}   
\end{figure}

We determine the Floquet multipliers $\mu(k)$ at $f=f_c=1.552$ 
numerically from the linear problem~(\ref{eq:forced-linear}). These 
are presented in Figure~\ref{fig:67hfm}. We find that the multipliers 
are well approximated away from the two primary resonance tongues by 
considering the unforced problem ($f=0$ in 
equation~\ref{eq:forced-linear}), for which 
\begin{equation}     
\label{eq:approxfm}   
\mu_\pm=e^{2 \pi \lambda_\pm}, \quad \lambda_\pm = -\gamma k^2 \pm i   
\Omega_k\ .     
\end{equation}       
Figures~\ref{fig:67hfm}a and \ref{fig:67hfm}b show the magnitude $\xi$ 
and the phase $\psi$ of the Floquet multipliers $\mu=\xi e^{i\psi}$ 
both as computed numerically from~(\ref{eq:forced-linear}) (solid 
line) and approximated by~(\ref{eq:approxfm}) (dotted 
line). Figure~\ref{fig:67hfm}c shows the real part of the Floquet 
multipliers, $\xi\cos \psi$, versus wavenumber $k$.  The ``bubbles" in 
this plot correspond to wavenumbers at which the Floquet multipliers 
are real (as opposed to a complex conjugate pair).  Weakly damped 
harmonic modes are associated with bubbles near a Floquet multiplier 
of $+1$.  Numerically we find that there are small bubbles of real 
Floquet multipliers whenever the phase $\psi$ is a multiple of $\pi$; 
this is demonstrated in Figure~\ref{fig:67hfm}d.  In particular, we 
find a bubble at wavenumber $k=0.383$, with associated real Floquet 
multiplier $\mu=0.93$. This mode is weakly damped and forms a resonant 
triad with primary harmonic modes separated by 
$\theta_r=22.2\degr$. (Here $k_{c,h}=0.997$ for the primary 
instability, which corresponds to $k_n$ in~(\ref{eq:theta-r-c}), with 
$k_m=0.393$ determined by the weakly damped harmonic mode.) Here we 
have focused on the wavenumbers associated with real Floquet 
multipliers near $\mu=+1$ since weakly damped modes with complex 
Floquet multipliers do not form a spatio--temporally resonant triad 
with the primary harmonic modes. 
   
We now present some hexagonal lattice bifurcation results for the 
specific parameters of this example, which are given in 
Figure~\ref{fig:linear}b. The computation of the quadratic and cubic 
coefficients in the bifurcation problem~(\ref{eq:bifurcation}) is 
described in the Appendix. 
We scale the amplitudes $z_j$ in~(\ref{eq:bifurcation}) so that 
$b_1=-1$, in which case we find that $\epsilon=0.00014$ and 
$b_2=-2.73$. Thus we expect results of Section~\ref{sec:bifurcation}, 
which focused on the unfolding of the degenerate bifurcation problem 
$\epsilon=0$, to apply. 
  
We find that simple hexagons, super hexagons and super triangles all 
bifurcate transcritically with the subcritical branch turning around 
in a saddle--node bifurcation. The stripes and rhombs solutions arise 
in supercritical pitchfork bifurcations. These claims are true for all 
lattice angles $\theta_h$ since the cubic coefficients 
$b_1,\ldots,b_6$ in~(\ref{eq:bifurcation}) are always negative; see 
Figure~\ref{fig:67nonlinear}a. Moreover, we find that simple hexagons 
are always stabilized in a saddle--node bifurcation and that they do 
not lose stability until after they reach the supercritical regime 
$\lambda>0$.  In contrast, super hexagons and super triangles are 
always unstable at $\lambda=0$, since at that point the sign of the 
second eigenvalue in Table~\ref{tab:evhex} is determined by 
$\sgn(b_1+2b_2-3b_4-3b_5-3b_6)$, which is positive for all $\theta$ 
(see Figure~\ref{fig:67nonlinear}b).  Thus, as $\lambda$ is increased 
through $0$, we expect a jump to finite amplitude simple hexagons as 
the other primary branches of~(\ref{eq:bifurcation}) are unstable. 
   
\begin{figure}     
\centerline{\resizebox{\textwidth}{!} {\includegraphics{67firstsh.eps}     
\hspace{0.6in} \includegraphics{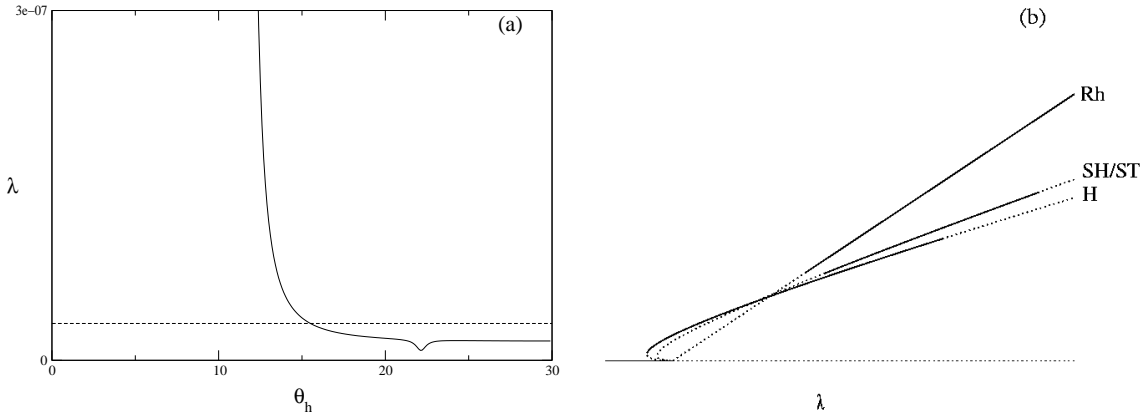}}}      
\caption{(a) $\lambda$ value at which the first (dashed) and second
(solid) eigenvalue expressions in~(\protect \ref{eq:simple}) turn
positive versus the lattice angle $\theta_h$.  Note that simple
hexagons (H) first lose stability to perturbations in the super
hexagon/triangle (SH/ST) direction at $\theta_h\approx 22.2\degr$.
(b) Schematic bifurcation diagram for the $(n_1,n_2)=(3,2)$ lattice of
Figure~\protect\ref{fig:hexlatt}, which corresponds to $\theta_h =
21.8\degr$. Stable (unstable) solutions are indicated by a solid
(dotted) line.  We do not show secondary branches or primary branches
that are never stable. The stable rhombs solution (Rh) corresponds to
one with an angle of $81.8\degr$, which is the rhombs solution closest
to $90\degr$ for this hexagonal lattice. The other two rhombs
solutions are unstable.}
\label{fig:67bifurcation}     
\end{figure}

We find that simple hexagons eventually lose stability as $\lambda$ 
increases since the following two expressions from 
Table~\ref{tab:evhex} change sign to positive (at least for some 
$\theta_h$) 
\begin{equation}   
\label{eq:simple} 
\sgn(-\epsilon x + (b_1-b_2)x^2),  
\quad \sgn(-\epsilon x + (b_4+b_5+b_6-b_1-2b_2)x^2).   
\end{equation}   
as the amplitude $x$ of simple hexagons grows with $\lambda$.  The 
first quantity changes from negative to positive at $\lambda 
\approx 3.2\times 10^{-8}$.  The second quantity changes sign with 
increasing $\lambda$ only for those values of $\theta_h$ where
$b_4+b_5+b_6-b_1-2b_2>0$, a condition which is met for $\theta_h \geq
11.5\degr$. Figure~\ref{fig:67bifurcation}a shows the value of
$\lambda$ where the expressions of~(\ref{eq:simple}) change
sign as a function of $\theta_h$. It follows that simple hexagons lose
stability {\it first} on the lattice with angle $\theta_h\approx
22.2\degr$.  This instability has an associated eigenvector in the
direction of super hexagon/triangles, and at this value of $\lambda$,
super hexagons (or triangles) are stable. These results are summarized
in Figure~\ref{fig:67bifurcation}b, which shows part of the
bifurcation diagram computed for the hexagonal lattice with
$(n_1,n_2)=(3,2)$, which corresponds to an angle
$\theta_h=21.8\degr$. Note that when simple hexagons lose stability,
both rhombs (Rh) and a superpattern are stable. Because the
instability that first destabilizes the simple hexagons is in the
direction of a superpattern with $\theta_h\approx 22.2\degr$, we
expect that the transition would be a hysteretic one involving the
simple hexagons and a superpattern, at least in the absence of noise
and other imperfections. We cannot determine whether the superpattern
is hexagonal or triangular from our calculations, since this requires
knowledge of fifth order terms in the bifurcation
problem~\cite{ref:dss97}.
 
\section{Conclusions}      
\label{sec:conclude}      
      
 
In this paper we have examined the effect of spatio-temporally 
resonant triads on two-dimensional pattern selection in parametrically 
excited systems. Using a normal form transformation to enforce temporal 
symmetry and center manifold reduction, we have argued that weakly 
damped \emph{harmonic} modes can strongly influence pattern selection 
by causing certain cubic cross-coupling coefficients in a 
twelve--dimensional ${\rm D}_6\dot{+}{\rm T}^2$--equivariant 
bifurcation problem to suddenly vary in magnitude for certain lattice 
angles~$\theta_h$. This suggests an important consideration in 
choosing one over another of the countable set of twelve--dimensional 
representations relevant to hexagonal bifurcation problems. Weakly 
damped \emph{subharmonic} modes, on the other hand, do not have such 
an effect. 
 
Our general analysis applies to any parametrically excited pattern 
forming system, but in particular is relevant to the interpretation of 
many recent experiments on two-frequency forced Faraday waves.  In 
such experiments, a bicritical point exists where subharmonic and 
harmonic instabilities are simultaneously excited.  On one side of the 
bicritical point, a subharmonic mode is excited and there is a weakly 
damped harmonic mode, while on the other, it is the harmonic mode 
which is excited and the subharmonic mode which is weakly damped. 
We showed that this weakly damped subharmonic mode does not influence
the harmonic wave pattern selection problem.  

We have derived the quadratic and cubic coefficients in the rhombic 
and hexagonal bifurcation equations describing the onset of patterns 
from the hydrodynamic equations of Zhang and Vi\~{n}als. We presented 
results for two different sets of parameters.  In the first case, the 
two forcing frequencies are in the ratio 2/3 and the modes near the 
bicritical point are the only ones of relevance.  As expected from our 
normal form analysis, for subharmonic waves, the weakly damped 
harmonic mode affects the cross--coupling coefficients, while 
for harmonic waves, the weakly damped subharmonic mode had no effect. 
 
In the second case of 6/7 forcing we have shown that, in addition to
the modes near the bicritical point, there are other harmonic modes
that are important.  These modes are not close to onset in the sense
that they only become critical at a much higher value of the
excitation amplitude, but {\em are} weakly damped and thus must be
taken into account. We demonstrate that they can have a stabilizing
effect on superlattice patterns at a lattice angle approximately equal
to the angle of the harmonic--harmonic resonance. This can occur if
the contribution of these weakly damped modes to the nonlinear
cross--coupling coefficient nearly cancels the other contributions to
this term, and hence is a subtle effect that depends on certain
details of the nonlinear problem, as well as the results of the linear
analysis which identifies the near critical modes. For the parameters
we have chosen, the onset pattern is simple hexagons, but upon a
further increase of the forcing, there is an instability to a
superlattice pattern associated with a hexagonal lattice with
$(n_1,n_2)=(3,2)$.
 
The experiments of Kudrolli, Pier and Gollub \cite{ref:kpg98} found a
superlattice pattern near the bicritical point which sits on a lattice
with $(n_1,n_2)=(3,2)$.  The work in this paper suggests that the
observation of this pattern could be explained by the interaction of
the primary harmonic instability and weakly damped harmonic modes.
However, the Zhang-Vi\~{n}als equations are not valid in the parameter
regime where this experiment was performed, and thus a study of the
full hydrodynamic problem is necessary to confirm this conjecture.  A
complete study should also involve a more complete analysis of the
codimension-2 bifurcation point and the associated dynamics, in the
spirit of J.D.~Crawford's early work on competing instabilities in the
Faraday problem~\cite{ref:c90a}. This would be of interest in light of
recent two--frequency experiments by Arbell and
Fineberg~\cite{ref:afpreprint} that show a variety of dynamic states
near the bicritical point, which involve both critical modes.
      
\appendix    
     
\label{sec:appendix}      
      
     
\section{Perturbation Theory}     
     
Here we outline the computation of the coefficients
in~(\ref{eq:rhombs}) and (\ref{eq:bifurcation}) from the equations of
Zhang and Vi\~{n}als (\ref{eq:ZVmodel}).  A multiple-scale perturbation
method is used to derive expressions for the coefficients which are
then evaluated numerically using a pseudospectral approach.  This
follows closely the method described in
\cite{ref:ss99} for the onset of one-dimensional patterns and 
we refer the reader there for further details. 
 
The coefficients can be derived by considering two different
calculations, namely the bifurcation problem (\ref{eq:bifurcation})
restricted in turn to a rhombic and a simple hexagons subspace.
     
\subsection{Rhombic lattice computation}     
     
In order to compute the coefficient $a$ and the cross-coupling
coefficient $\beta(\theta)$ in~(\ref{eq:rhombs}) we seek solutions
which are periodic on a rhombic lattice associated with an angle
$\theta$.  We are thereby able to compute the coefficients $b_1$,
$b_4$, $b_5$, and $b_6$ in the bifurcation equations
(\ref{eq:bifurcation}) since $b_1=a$, $b_4=\beta(\theta_h)$,
$b_5=\beta(\theta_h+2\pi/3)$, and $b_6=\beta(\theta_h-2\pi/3)$.
 
First we    
introduce a small parameter $\eta$, such that    
\begin{eqnarray}     
h(x,y,\tau)&=&\eta h_1(x,y,\tau,T)+\eta^2 h_2(x,y,\tau,T) \\     
& & \mbox{}+\eta^3 h_3(x,y,\tau,T)+\cdots \nonumber \\      
\Phi(x,y,\tau)&=&\eta \Phi_1(x,y,\tau,T)+\eta^2  \Phi_2(x,y,\tau,T) \nonumber \\     
&&\mbox{}+\eta^3 \Phi_3(x,y,\tau,T)+\cdots, \nonumber      
\end{eqnarray}      
in (\ref{eq:ZVmodel}) where      
\begin{equation}      
T=\eta^2\tau , \quad f=f_{c}+\eta^2 f_2.      
\end{equation}      
Here $f_{c}$ is the critical excitation amplitude. 
The terms in the expansion for $h$ and $\Phi$  
may be written in the following separable Floquet-Fourier    
form:    
\begin{eqnarray}      
h_1&=&[w_{1}(T)e^{ik_{c}x}+w_{4}(T)e^{ik_{c}(cx+sy)}+c.c.]p_1(\tau) \\      
\Phi_{1}&=&[w_{1}(T)e^{ik_{c}x}+w_{4}(T)e^{ik_{c}(cx+sy)}+c.c.]q_1(\tau)     
\nonumber \\      
h_2&=&[w_{1}^{2}(T)e^{2ik_{c}x}+w_{4}^{2}(T)e^{2ik_{c}(cx+sy)}]p_{2,1}(\tau)     
\nonumber \\         
&&\mbox{}+w_{1}(T)\overline{w}_{4}(T)e^{ik_{c}((1-c)x-sy)}p_{2,2}(\tau) \nonumber \\        
 &&\mbox{}+w_{1}(T)w_{4}(T)e^{ik_{c}((1+c)x+sy)}p_{2,3}(\tau)+c.c. \nonumber \\      
\Phi_2&=&[w_{1}^{2}(T)e^{2ik_{c}x}+w_{4}^{2}(T)e^{2ik_{c}(cx+sy)}]q_{2,1}(\tau)     
\nonumber \\     
&&\mbox{}+w_{1}(T)\overline{w}_{4}(T)e^{ik_{c}((1-c)x-sy)}q_{2,2}(\tau) \nonumber \\      
&&\mbox{}+w_{1}(T)w_{4}(T)e^{ik_{c}((1+c)x+sy)}q_{2,3}(\tau)+c.c. \nonumber      
\end{eqnarray}      
where $c=\cos \theta$, $s=\sin \theta$, and $\theta$ is not a multiple    
of $\frac{\pi}{3}$. Here $p_1$ and $q_1$ are real $2\pi$-periodic    
functions of the fast time $\tau$ in the case of harmonic waves; in    
the case of subharmonic waves they are $4\pi$-periodic in $\tau$.    
Additionally, $p_{2,r}$ and $q_{2,r}$ ($r=1,2,3$) are real    
$2\pi$-periodic functions of $\tau$.  The wave number $k_{c}$ is    
associated with the onset unstable mode.    
     
At $\mathcal{O}(\eta)$ we recover the linear problem which determines    
$k_{c}$ and $f_c$, as well as the functions $p_1$, $q_1$    
to within a multiplicative constant.  At $\mathcal{O}(\eta^2)$,  
equations are found which allow us to solve for the  
functions $p_{2,r}$ and $q_{2,r}$.   
Finally, at    
$\mathcal{O}(\eta^3)$, we apply a solvability condition to ensure that    
a periodic solution exists.  This condition leads to the amplitude    
equations    
\begin{eqnarray}     
\label{eq:rhombic-landau}     
\delta{d w_1 \over d T} & = & \alpha f_2 w_1+A|w_1|^2w_1 +B(\theta)|w_4|^2w_1 \\     
\delta{d w_4 \over d T} & = & \alpha f_2 w_4+A|w_4|^2w_4 +B(\theta)|w_1|^2w_4,      
\nonumber      
\end{eqnarray}      
where      
\begin{eqnarray}      
\label{eq:rhombcoeffs}     
\delta & = & { 1\over 2\pi} \int_0^{4\pi} (p_1' + \gamma      
k_{c}^2p_1) \widetilde p_1\ d\tau \\     
\alpha & = & {k_{c}\over 4\pi}\int_0^{4\pi} [\cos(\chi)\cos(m\tau)+\sin(\chi)      
\cos(n\tau+\phi)]p_1\widetilde p_1\ d\tau \nonumber \\     
A & = & {k_{c}^2 \over 4 \pi} \int_0^{4\pi} \bigl[-k_{c}(p_1^2 q_1)' -\gamma k_{c}^3     
p_1^2 q_1 - 2(q_1 p_{2,1})' - 2 \gamma k_{c}^2 q_1 p_{2,1} \nonumber \\     
& & \quad \quad \quad \mbox{}+k_{c}^2 q_1^2 p_1 + {3\over     
2} k_{c}^3 \Gamma_0 p_1^3 \bigr] \widetilde p_1 \  d\tau \nonumber \\      
B(\theta) & = & {k_{c}^2\over 4\pi}\int_0^{4\pi}     
\Bigl[(1-c-\sqrt{2-2c})[(p_1q_{2,2})' + \gamma k_{c}^2  p_1q_{2,2} - k_{c}q_1 q_{2,2}] \nonumber \\     
&&\quad \quad \mbox{}+(1+c-\sqrt{2+2c})[(p_1q_{2,3})' +\gamma k_{c}^2 p_1q_{2,3} - k_{c}q_1 q_{2,3}] \nonumber \\      
&&\quad \quad \mbox{}-(1-c)[(p_{2,2}q_1)' + \gamma k_{c}^2  p_{2,2}q_1]  
-(1+c)[(p_{2,3}q_1)'+\gamma k_{c}^2 p_{2,3}q_1] \nonumber \\      
&&\quad \quad \mbox{}-(6-2\sqrt{2-2c}-2\sqrt{2+2c})[k_c(p_1^2q_1)'  
+\gamma k_{c}^3p_1^2q_1 - k_{c}^2p_1q_1^2] \nonumber \\     
&&\quad \quad \mbox{}+\Gamma_{0}(3c^2+s^2)k_{c}^3p_1^3)\Bigr]    
\widetilde p_1\ d\tau. \nonumber     
\end{eqnarray}     
In the above, a prime denotes differentiation with respect to $\tau$  
and $\widetilde p_1$ is the equivalent of $p_1$ for the adjoint 
problem at $\mathcal{O}(\eta)$.  
%
%
%
The amplitude equations (\ref{eq:rhombic-landau}) may be re-scaled and 
then comparison with the map (\ref{eq:rhombs}) yields 
\begin{equation}     
a= b_1 = \sgn (A \alpha),  
    \quad \beta(\theta) = \sgn (A \alpha) \frac{B(\theta)}{A}.     
\end{equation}     
 
     
\subsection{Hexagonal lattice computation}     
     
Similarly, we compute the coefficients $\epsilon$ and $b_2$     
in the bifurcation equations (\ref{eq:bifurcation}) by  
seeking solutions in the form of simple     
hexagons.  Here we use a three-timing perturbation method,  
writing the solution as     
\begin{eqnarray}     
h(x,y,\tau)&=&\eta h_1(x,y,\tau,T_1,T_2)+\eta^2     
h_2(x,y,\tau,T_1,T_2) \\     
&&\mbox{}+\eta^3 h_3(x,y,\tau,T_1,T_2)+\cdots \nonumber \\      
\Phi(x,y,\tau)&=&\eta \Phi_1(x,y,\tau,T_1,T_2)+\eta^2      
\Phi_2(x,y,\tau,T_1,T_2) \nonumber \\      
&&\mbox{}+\eta^3 \Phi_3(x,y,\tau,T_1,T_2)+\cdots, \nonumber      
\end{eqnarray}      
where      
\begin{equation}      
T_1=\eta\tau , \quad T_2=\eta^2\tau,      
\end{equation}      
and
\begin{eqnarray}      
h_1&=&w_{1}(T_1,T_2)p_{1}(\tau)[e^{ik_{c}x}+e^{ik_{c}(-{1 \over     
2}x+{\sqrt{3} \over 2 }y)}+e^{ik_{c}(-{1 \over 2}x-{\sqrt{3} \over 2}y)}+c.c.] \\     
\Phi_1&=&w_{1}(T_1,T_2)q_{1}(\tau)[e^{ik_{c}x}+e^{ik_{c}(-{1 \over     
2}x+{\sqrt{3} \over 2 }y)}+e^{ik_{c}(-{1 \over 2}x-{\sqrt{3} \over 2}y)}+c.c.] \nonumber \\     
h_2&=&w_1^2(T_1,T_2)\bigl\{     
p_{2,1}(\tau)[e^{ik_{c}2x}+e^{ik_{c}(-x+\sqrt{3}y)}+e^{ik_{c}(-x-\sqrt{3}y)}+c.c.] \nonumber \\     
&&\mbox{}+p_{2,2}(\tau)[e^{ik_{c}x}+e^{ik_{c}(-{1 \over 2}x+{\sqrt{3} \over     
2 }y)}+e^{ik_{c}(-{1 \over2}x-{\sqrt{3} \over 2 }y)}+c.c.] \nonumber \\     
&&\mbox{}+p_{2,3}(\tau)[e^{ik_{c}({3 \over 2}x-{\sqrt{3} \over 2     
}y)}+e^{ik_{c}\sqrt{3}y}+e^{ik_{c}({3 \over 2}x+{\sqrt{3} \over 2}y)}+c.c.] \bigr\}     
\nonumber \\     
\Phi_2&=&w_1^2(T_1,T_2)\bigl\{     
q_{2,1}(\tau)[e^{ik_{c}2x}+e^{ik_{c}(-x+\sqrt{3}y)}+e^{ik_{c}(-x-\sqrt{3}y)}+c.c.] \nonumber \\     
&&\mbox{}+q_{2,2}(\tau)[e^{ik_{c}x}+e^{ik_{c}(-{1 \over 2}x+{\sqrt{3} \over     
2 }y)}+e^{ik_{c}(-{1 \over2}x-{\sqrt{3} \over 2 }y)}+c.c.] \nonumber \\     
&&\mbox{}+q_{2,3}(\tau)[e^{ik_{c}({3 \over 2}x-{\sqrt{3} \over 2     
}y)}+e^{ik_{c}\sqrt{3}y}+e^{ik_{c}({3 \over 2}x+{\sqrt{3} \over 2}y)}+c.c.] \bigr\}.     
\nonumber     
\end{eqnarray}     
As with the rhombic case, $p_1$, $q_1$, $p_{2,r}$ and $q_{2,r}$ are real.  Additionally, we take the     
amplitude $w_1(T_1,T_2)$ to be real.     
     
For the harmonic case, at $\mathcal{O}(\eta^2)$  
the  solvability condition, 
\begin{equation}     
\label{eq:hex1}     
\delta \frac{\partial w_1}{\partial T_1} = \beta_0 w_1^2,  
\end{equation}     
must be satisfied,   
where $\delta$ is given by (\ref{eq:rhombcoeffs}).  The quadratic  
coefficient  is     
\begin{equation}     
\beta_0 = \frac{k_{c}^2}{4 \pi} \int_0^{4 \pi} [-(p_1 q_1)' - \gamma k_{c}^2 p_1     
q_1 + {1 \over 2} k_{c} q_1^2] \widetilde{p_1}\ d\tau\ .     
\end{equation}     
There is no solvability condition for subharmonic waves at 
$\mathcal{O}(\eta^2)$, reflecting the fact that there are 
no even terms in the amplitude equations (\ref{eq:bifurcation}) 
for this case. 
     
At  $\mathcal{O}(\eta^3)$, we again apply a solvability condition to     
ensure that a periodic solution exists.  This conditions leads to the     
amplitude equation     
\begin{equation}     
\label{eq:hex2}     
\delta \frac{\partial w_1}{\partial T_2} 
= \alpha f_2 w_1 + (A+2\beta_2) w_1^3     
\end{equation}     
The coefficients $\delta$, $\alpha$, and $A$ are given by (\ref{eq:rhombcoeffs}), and     
\begin{eqnarray}     
\beta_2 & = & \frac{1}{4 \pi} \int_0^{4 \pi}     
\Bigl[ (\frac{3}{2}-\sqrt{3})k_c^2[(p_1 q_{2,3})'  
     + \gamma k_{c}^2 p_1 q_{2,3} - k_{c} q_1 q_{2,3}]\\     
&&\quad \quad \mbox{} + (2 \sqrt{3}-4)k_{c}^3[(p_1^2 q_1)' 
     +\gamma k_{c}^2 p_1^2 q_1 - k_{c} p_1 q_1^2] \nonumber \\     
&&\quad \quad \mbox{} - \frac{3}{2}k_{c}^2[(p_{2,3}q_1)' 
     +\gamma k_{c}^2 p_{2,3} q_1 - \Gamma_0 k_{c}^3 p_1^3] \nonumber \\      
&&\quad \quad \mbox{} - \frac{1}{2}k_{c}^2 [(p_1 q_{2,2})'  
     + \gamma k_{c}^2 p_1 q_{2,2} + (p_{2,2} q_1)'  
+ \gamma k_{c}^2 p_{2,2} q_1 - k_{c} q_1 q_{2,2}]  \nonumber \\     
&&\quad \quad \mbox{} - \frac{\beta_0}{\delta}[k_{c}^2 p_1 q_1 + \frac{\beta_0}{\delta}p_1      
+ 2 p_{2,2}' + 2 \gamma k_{c}^2 p_{2,2}] \Bigr] \widetilde p_1 \ d\tau. \nonumber     
\end{eqnarray}     
By rescaling $\eta w_j(T_1,T_2) \rightarrow w_j(T)$ and $\eta^2     
\alpha f_2 \rightarrow \alpha f_2$, we obtain the reconstituted     
hexagonal bifurcation equation     
\begin{equation}     
\label{eq:hex-landau}     
\delta \frac{d w_1}{d T} = \alpha f_2 q_1 + \beta_0 w_1^2 + (A + 2\beta_2) w_1^3     
\end{equation}     
Finally, after  
rescaling as for the rhombic case, and  
comparing (\ref{eq:hex-landau}) to (\ref{eq:bifurcation})  
we find that     
\begin{equation}     
\epsilon = \sgn (\alpha)\frac{\beta_0}{\sqrt{|\alpha A|}},  
\quad b_2 = \sgn (A \alpha) \frac{\beta_2}{A}.     
\end{equation}     
     
\section*{Acknowledgments}      

First and foremost, we acknowledge the tremendous influence of John
David Crawford's research on parametrically excited wave patterns on
our own study of Faraday waves. MS also wishes to acknowledge the
inspiration she has always taken from John David's work, as well as
the encouragement she received from him.

The research of MS was supported by NSF grant DMS-9972059
and by an NSF CAREER award DMS-9502266.
     
\bibliographystyle{unsrt}     
\bibliography{revisedsts.bib}     
     
\end{document}